\documentclass[aps,prc,twocolumn]{revtex4}
\usepackage{graphicx}
\usepackage{amsmath,bbm}
\usepackage{amssymb,bm}
\usepackage{array}
\usepackage{multirow}
\usepackage{caption}
\usepackage{subcaption}
\newcommand\nd{\noindent}
\newcommand \sH {\mathcal{H}}
\newcommand \sL {\mathcal{L}}
\newcommand \sK {\mathcal{K}}

\newcommand \bx {{\bf{x} }}
\newcommand \bp {{\bf{p} }}

\newcommand \bigO {\mathcal{O}}
\newcommand \sS {\mathcal{S}}

\begin{document}
\title{Many body gravity and the galaxy rotation curves} 
\author{S.~Ganesh\footnote{Corresponding author:\\Email: gans.phy@gmail.com}}
\affiliation{Independent Researcher}
\begin{abstract}

	A novel theory was proposed earlier to model systems with thermal gradients, based on the postulate that the spatial and temporal variation in temperature can be recast as a variation in the metric.
	Combining the variation in the metric due to the thermal variations and gravity, leads to the concept of thermal gravity in a 5-D space-time-temperature setting. When the 5-D Einstein field equations are projected to a 4-D space, they result in additional terms in the field equations. This may lead to unique phenomena such as the spontaneous symmetry breaking of scalar particles in the presence of a strong gravitational field. 
	This theory, originally conceived in a quantum mechanical framework, is now adapted to explain the galaxy rotation curves.
	A galaxy is not in a state of thermal equilibrium. A parameter called the "degree of thermalization" is introduced to model partially thermalized systems. The generalization of thermal gravity to partially thermalized systems, leads to the theory of many-body gravity. The theory of many-body gravity is now shown to be able to explain the rotation curves of the Milky Way and the M31 (Andromeda) galaxies, to a fair extent. 
	The radial acceleration relation (RAR) for 63 galaxies, with their galactic masses spanning three orders of magnitude, has been replicated. Finally, the wide binary star (WBS) system is touched upon.

\vskip 0.5cm

{\nd \it Keywords } : Many body gravity , RAR, Galaxy rotation curves, Thermal gravity, 5-D thermal field theory \\
{\nd \it PACS numbers } : 04.50.Kd, 98.62.Dm, 11.10.Wx, 

\end{abstract}

\maketitle
\section{Introduction}
\label{sec:intro}
	Einstein's field equations~\cite{Ein1} have proven to be very effective in explaining the phenomenon of gravity. Various experiments~\cite{shap1, hol, sch, foma, shap2,ber}
	including the detection of gravitational waves~\cite{ligo1, ligo2} and gravitational lensing~\cite{lens1, lens2} have reinforced Einstein's general theory of relativity.
	The galaxy rotation curves have however been baffling, to say the least. Newtonian gravity, or Einstein's general theory of relativity, predicts the orbital speeds of stars around the center of the galaxy to decrease radially, as one goes away from the center. However, the stars are seen to have a constant speed~\cite{rub1, rub2, galaxy1, galaxy2}.
	Many theories have been formulated to explain the rotational speed of stars or what is known in the literature as the galaxy rotation curves.
	The theories propounded to explain the galaxy rotation curves can be classified into two categories. The first category hypothesizes the existence of a hidden mass or dark matter~\cite{dark1, dark2, dark3, dark4, dark5}, while the second looks at the modification of the laws of gravity, especially for weak gravitational fields~\cite{mond1, mond2, mond3, mond4, mond5, entropic}.
	The current article falls within the category of modified gravity. However, importantly, the theory has not been developed with the goal of explaining galaxy rotation curves, unlike many other such theories. The genesis of the current theory lies in modeling spatial and temporal variations of thermal systems, and can be validated experimentally~\cite{gans6,gans7}. Moreover, the theory is manifestly relativistic.

	Reference~\cite{gans5} first introduced the concept of recasting the variation in the temperature as a variation in the metric. This was done for systems in local thermal equilibrium. The concept was used to calculate the quark-antiquark potential, in a Quark Gluon Plasma, using the Anti De-Sitter/Conformal Field Theory (AdS-CFT) correspondence. The concept was placed on firmer grounds in Ref.~\cite{gans6}, using the Polyakov loop, the partition function, the geodesic equation etc. Calculations were performed in the field theoretic domain to calculate the pressure and the energy density of a scalar gas system, possessing spatial thermal gradients. 
	The interpretation of the geodesic equation in the context of a curved space due to thermal variations was explained in Ref.~\cite{gans6}.
	The formulation was extended to thermal systems with temporal variations in a 5-D space-time-temperature framework, in Ref.~\cite{gans7}.  Calculations were made for the energy density of a scalar gas system with a time varying temperature. Gravitational fields were combined with the metric modifications for temperature variations. This was shown to lead to novel phenomena like the spontaneous symmetry breaking of scalar fields under a very strong gravitational field, if the scalar field is non-minimally coupled with the Ricci scalar. The new phenomenon manifests due to the additional terms that appear in the Ricci tensor, if the 5-D Einstein field equations are expressed in terms of 4-D operators~\cite{gans7}. 

	The above formulation is now applied to explain the galaxy rotation curves. Unfortunately, the application is not straightforward, as a galaxy does not constitute a thermal medium in thermal equilibrium.
	However, a system being or not being in thermal equilibrium is not a binary state. There could be varying degrees of equilibration. It would depend on the extent to which the particles constituting the system have interacted and influenced each other.
	A system of particles with nil interaction with each other, continue to be in a state of absolute non-equilibrium. The behavior of one particle has no bearing on another particle, and the system is just a collection of individual particles.
	A system, where particles interact with each other to such an extent, that only the ensemble properties describe the state of the system (namely the Fermi-Dirac or Bose-Einstein statistics), can be said to be in thermal equilibrium.
	While the stars in a galaxy are not in thermal equilibrium (either local or global), they do have a certain degree of interaction with each other. A given star would be affected by the tugs and pulls of the nearby stars and the interstellar gasses.
	To capture the aforementioned phenomenon, a degree of equilibration parameter, $k$, is introduced. 
	A system with $k=0$, constitutes a completely non-interacting system, and thus in complete non-equilibrium. In this case, we shall later see that the equations of gravity reduce to the classical Einstein field equations in a 4-D space-time. 

	The rest of the article is as follows. 
	An overview of the theory developed in Refs.~\cite{gans6,gans7} is given in Sec.~\ref{sec:overview}. The application of the theory to galaxy rotation curves begins from Sec.~\ref{sec:formulation}.
	The field equations in 5-D space-time-temperature are solved in the weak field limit in Sec.~\ref{sec:formulation}. Numerical simulations are performed in Sec.~\ref{sec:numerical} for Milkyway and M31. 
	Other phenomena such as pseudo mass, RAR, wide binary star systems, the Bullet cluster and the ultra diffuse galaxies (UDG) are touched upon in Sec.~\ref{sec:prediction}.
	Finally, Sec.~\ref{sec:conclusion} draws the conclusions.

\section{An overview of the underlying theory}
\label{sec:overview}
	A brief overview of the theory developed in Refs.~\cite{gans6,gans7} is now provided. The reader is, however, referred to Refs.~\cite{gans6,gans7}, for the motivation, derivations, inferences and details of the theory.

\subsection{The 5-D and 8-D space}
\label{sec:5D8D}
	We begin by summarizing the 5-D and 8-D spaces, introduced in Ref.~\cite{gans6}.
A covariant approach could involve considering a complex 4-D space as $X^{\mu} \equiv x^{\mu} + i\beta^{\mu}$, with $x^{\mu}$ being the conventional space-time, and $\beta^{\mu} = \beta u^{\mu}$, where, $u^{\mu}$ is the four-velocity representing the flow of the thermal medium, and $\beta$ is the inverse temperature.
It is trivially seen that this space, under a Lorentz boost, would transform as, $X'^{\nu} = \Lambda^{\nu}_{\mu} X^{\mu}$, where $\Lambda^{\nu}_{\mu}$ represents the Lorentz transformation.
If the thermal medium is stationary, i.e., $u^{\mu} = (1,0,0,0)$, then the complex space-time reduces to $(t + i\beta, x, y, z)$.
In the case of a time invariant system, it is sufficient to consider the sub-space $(i\beta, x, y, z)$, which forms the co-ordinate space for the imaginary time formalism.
One might also be tempted to view the original 4-D complex space, $x^{\mu} + i\beta^{\mu}$, as a 8-D, $\beta^{\mu}\times x^{\nu}$, space.
The 8-D space under a Lorentz boost would transform as:
                \begin{equation}
                \label{eq:8D}
                \left[ \begin{array}{c}
                        \beta'^{\gamma}\\
                        x'^{\delta}\\
                \end{array} \right ]
                         =
                \left[ \begin{array}{c c}
                        \Lambda^{\gamma}_{\alpha}&0\\
                        0&\Lambda^{\delta}_{\beta}\\
                \end{array} \right ]
                \left[ \begin{array}{c}
                        \beta^{\alpha}\\
                        x^{\beta}\\
                \end{array} \right ].
                \end{equation}
Let us consider the following metric tensor in this 8D space:
                \begin{equation}
                        \label{eq:8Dmetric}
                        G^{8D} =
                \left[ \begin{array}{c c}
                        h_{\rho \sigma}&0\\
                        0&g_{\mu\nu}\\
                \end{array} \right ].
                \end{equation}
The $\rho,\sigma,\mu$ and $\nu$ indices vary within their 4-D subspaces respectively.
The metric elements, $g_{\mu\nu}$, are the usual metric elements describing the curvature of space-time, $x^{\mu}$, under a gravitational field. 
In the limit $u^{\mu} = (1,0,0,0)$,
the 8-D $\beta^{\mu} \times x^{\nu}$ space,
reduces to the five dimensional space $(i\beta, t, x, y, z)$.
Under the additional more restrictive conditions,
\begin{itemize}
\item the thermal bath and the system immersed in it is time invariant,
\item  $g_{00} = 1$; $g_{0i} = 0~\forall i = 1,2,3$,
\end{itemize}
it suffices to consider the 4-D Euclidean subspace, $(i\beta,x,y,z)$, which again leads to the imaginary time formalism.
The metric for the 4-D sub-space, $(i\beta, x, y, z)$, can be inferred from Eq.~\ref{eq:8Dmetric} as:
                \begin{equation}
                        \label{eq:4Deuc_metric}
                        G^{Euc} =
                \left[ \begin{array}{c c}
                        h_{00}&0\\
                        0&g_{ij}\\
                \end{array} \right ],
                \end{equation}
where $i,j = 1,2,3$.
The interpretation of $h_{00}$ is explained in Sec.~\ref{sec:partition}.

\subsection{The partition function}
\label{sec:partition}
In Ref.~\cite{gans6}, the Quantum field theoretic aspects of a non-time varying thermal field with spatial variations, were developed. In Ref.~\cite{gans7}, it was extended to time varying thermal fields in the 5-D space-time-temperature $(i\beta, t, x,y,z)$.
We now summarize the Quantum field theoretic aspects of the more generic case of a time varying thermal field, developed in Ref.~\cite{gans7}. Importantly, this section provides a brief as to why the variation in the temperature can be recast as a variation in the metric~\cite{gans6,gans7}.
This would form a crucial aspect of the theory, which would subsequently be used to model the galaxy rotation curves.
The field theory is developed for a non-interacting Lagrangian, in 5-D space-time-temperature. The Lagrangian density in a 5-D space, for a neutral scalar field would be,
\begin{equation}
\label{eq:lagrangian5D}
        \sL(\hat{\phi},\partial_a \hat{\phi}) =  \frac{1}{2} \left ( \partial_{a}\hat{\phi} \partial^{a}\hat{\phi} - m^2\hat{\phi}^2 \right ),
\end{equation}
with the index $a=0,1,2,3,4$ corresponding to the dimensions $[\tau,t,x,y,z]$.
$\tau$ is the inverse temperature and varies from $0$ to $\beta$.
The sign convention used is (-,+,-,-,-).
This gives rise to the equation of motion:
\begin{equation}
        \partial_a\partial^a\hat{\phi} + m^2\hat{\phi} = 0.
\end{equation}
The constraining 5 momentum delta function would be, $\delta( E^2 - \omega^2 - \bp^2 - m^2)$, and the 5-D integral measure is:
\begin{multline}
        \label{eq:invariant}
        \frac{1}{\beta}\sum_{n}\int \frac{d^3p}{(2\pi)^3}\frac{dE}{2\pi} 2\pi\delta(E^2 - \omega_n^2 - \bp^2 -m^2 )\\
        =  \frac{1}{\beta}\sum_{n}\int \frac{d^3p}{(2\pi)^3} \frac{1}{2E},
\end{multline}
where, the Matsubara frequency, $\omega_n = \frac{2n\pi}{\beta}$, for a boson.
As mentioned in Ref.~\cite{gans6}, $E$, may be considered the original intrinsic energy of a particle, and $\omega_n = iE_c$, can be considered as the interaction energy of the particle with the thermal medium. The variable, $\omega_n$, determines the decay or enhancement of a particle wave-function with temperature (for example, the Dirac spinor in Ref.~\cite{gans6}).
Since $E$ and $\omega_n$ lie in orthogonal dimensions (conjugate momenta to time and temperature respectively), the magnitude of the total energy is then $= \sqrt{E^2 - \omega_n^2} (= \sqrt{E^2 + E_c^2}) $. It is intuitive, that a particle's 3-momentum would be affected by both $E$ and $E_c$, and not just $E$. Thus, one may consider $E^2 - \omega_n^2 = \bp^2 + m^2$. A portion of the particle's original energy, $E$, is lost due to interaction with the thermal medium. This provides an intuition behind the delta function, $\delta(E^2 - \omega_n^2 - \bp^2 - m^2)$.

The operator for a neutral scalar field in 5-D space-time-temperature is,
\begin{multline}
        \label{eq:operator5D}
        \hat{\phi}(\bx,\tau,t) = \frac{1}{\beta}\sum_n \int \frac{d^3p}{(2\pi)^3} \frac{1}{\sqrt{2E_p}}\\
        \sum_s \left ( a^{\dagger}_{\bp,\omega_n} e^{-ipx}e^{-i\omega_n \tau} + a_{\bp,\omega_n}e^{ipx}e^{i\omega_n \tau} \right ).
\end{multline}

The operator, $a^{\dagger}_{\bp,\omega_n}$, creates a particle with 3-momentum $\bp$, and Matsubara frequency $\omega_n$.

One may premise the below commutation relation:
\begin{equation}
        \label{eq:commutation5D}
        [a_{\bp_1,\omega_{n1}}, a^{\dagger}_{\bp_2,\omega_{n2}}] = (2\pi)^3\delta^3(\bp_1 - \bp_2) \zeta(\beta)\delta_{n1,n2},
\end{equation}
where, $\zeta(\beta)$ is a scalar normalization function, and needs to be determined.
Let us define,
\begin{eqnarray}
        \label{eq:pcreation}
        \nonumber       a_{\bp} = \sum_n f(\omega_n) a_{\bp,\omega_n}, \\
        a^{\dagger}_{\bp} = \sum_n f^*(\omega_n) a^{\dagger}_{\bp,\omega_n}.
\end{eqnarray}
Eq.~\ref{eq:pcreation} can be interpreted in the following way. When a momentum state $|\bp\rangle$ is created, then $|\bp\rangle$ itself can be treated as a superposition of the momentum-Matsubara eigenstates $|\bp,\omega_n\rangle$, with probability amplitudes $f(\omega_n)$.
Since $f(\omega_n)$ is a probability amplitude, $\sum_n|f(\omega_n)|^2 = 1$.
In Ref.~\cite{gans7}, it was shown that the commutator,
\begin{multline}
\label{eq:equaltau}
[a_{\bp_1}, a^{\dagger}_{\bp_2}] 
        = \sum_{n1} (2\pi)^3 \zeta(\beta) \delta^3(\bp_1 - \bp_2)|f(\omega_{n1})|^2.
\end{multline}
Since $\sum_{n1} |f(\omega_{n1})|^2 = 1$,  let us assign $\zeta(\beta) =  1$, in Eq.~\ref{eq:equaltau}, to obtain,
\begin{equation}
        \label{eq:commutator4D}
        [a_{\bp_1}, a^{\dagger}_{\bp_2}] = (2\pi)^3\delta^3(\bp_1 - \bp_2).
\end{equation}
Thus, the usual commutation relation between the 3-momentum annihilation and creation operator is recovered.

We now proceed to determine the conjugate momenta and the Hamiltonian.
There can be a conjugate momenta w.r.t. either the time variable or the temperature variable, i.e.,
\begin{equation}
        \label{eq:pi_timetempaxis}
        \hat{\pi}_t = \frac{\delta \sL}{\delta \frac{\partial \hat{\phi}}{\partial t}};~~~
        \hat{\pi}_{\beta} = i\frac{\delta \sL}{\delta \frac{\partial \hat{\phi}}{\partial \tau}}.
\end{equation}
The corresponding Hamiltonian densities are:
\begin{equation}
        \label{eq:hamil_bothaxis}
        \sH_t = \hat{\pi}_t \frac{\partial \hat{\phi}}{\partial t} - \sL;~~
        \sH_{\beta} = -i\hat{\pi}_{\beta} \frac{\partial \hat{\phi}}{\partial \tau} - \sL.
\end{equation}

They would obey the evolution equations:
\begin{equation}
        i\frac{\partial \hat{\phi}}{\partial t} = [\hat{\phi},H_t];~~~
        \frac{\partial \hat{\phi}}{\partial \tau} = [\hat{\phi},H_{\beta}].
\end{equation}

Since we are modeling a thermal system, and are interested in evolution in $\tau$, the main object of interest would be $H_{\beta}$ and $\pi_{\beta}$. For convenience, we now drop the subscript $\beta$. In the rest of this section, unless otherwise mentioned, $H$ and $\hat{\pi}$ refer to $H_{\beta}$ and $\hat{\pi}_{\beta}$.
We now follow a similar procedure as Ref~\cite{gans6,kapusta}, albeit modified for a 5-D space with thermal variations.
Let $\phi(\bx)$ and $|\phi(\bx)\rangle$ be the eigenfunction and the eigenket of the Schrodinger picture field operator $\hat{\phi}(\bx,0,0)$, while, $\pi(\bx)$ and $|\pi(\bx)\rangle$ be the eigenfunction and the eigenket of the conjugate momentum field operator $\hat{\pi}(\bx,0,0)$. In other words,
\begin{eqnarray}
\nonumber       \hat{\phi}(\bx,0,0)|\phi\rangle = \phi(\bx)|\phi\rangle,\\
        \hat{\pi}(\bx,0,0)|\pi\rangle = \pi(\bx)|\pi\rangle.
\end{eqnarray}
The eigenkets, $|\phi \rangle$ and $|\pi \rangle$, obey the following relation:
\begin{equation}
        \langle \phi| \pi \rangle = \exp\left (i\int d^3x \pi(\bx)\phi(\bx) \right ).
\end{equation}
For a time-dependent system, the time-dependent Hamiltonian can be written as:
\begin{equation}
        H(t) = H_0 + H_I(t),
\end{equation}
where $H_0$ is the time independent part, and $H_I(t)$ be the time dependent part.
$H_0$ is written in terms of the Schrodinger picture operators $\hat{\pi}(\bx,0,0)$ and  $\hat{\phi}(\bx,0,0)$.
\begin{equation}
        H_0 = \int d^3x \sH_0(\hat{\pi}(\bx,0,0), \hat{\phi}(\bx,0,0)) \equiv \int d^3x \sH_0(\bx).
\end{equation}
For simplicity of notation, the form, $\sH_0(\bx)$, is used as a representation of $\sH_0(\hat{\pi}(\bx,0,0), \hat{\phi}(\bx,0,0))$.

The free Hamiltonian density, corresponding to the Lagrangian in Eq.~\ref{eq:lagrangian5D}, would then be,
\begin{equation}
        \label{eq:Hamiltonian5D}
                \sH_0 = \frac{1}{2}\left ( \hat{\pi}^2 + (\nabla \hat{\phi})^2 - \left ( \frac{\partial \hat{\phi}}{\partial t} \right )^2 + m^2\hat{\phi}^2 \right ).
\end{equation}
But, $\frac{\partial \hat{\phi}}{\partial t} = -iE\hat{\phi}$. In a gas composed of scalar fields, which is equilibrated, $E \rightarrow 0$, as only the ensemble interaction energy, captured by the Matsubara frequency, $\omega_n$, is non zero~\cite{gans6}.
We are then left with the standard imaginary time formalism.
On similar lines, in a vacuum, as $\beta \rightarrow \infty$, $\omega_n \rightarrow 0$. Then, the only energy left is the particle energy, $E$. The Lagrangian in Eq.~\ref{eq:lagrangian5D}, then boils down to the normal 4-D space-time Quantum Field Theory (QFT).
With, $E\rightarrow 0$, the Hamiltonian density in Eq.~\ref{eq:Hamiltonian5D} becomes,
\begin{equation}
        \label{eq:hamiltonian4D}
        \sH_0 = \frac{1}{2}\left ( \hat{\pi}^2 + (\nabla\hat{\phi})^2  + m^2\hat{\phi}^2 \right ).
\end{equation}
Equations \ref{eq:commutator4D} and \ref{eq:hamiltonian4D},
indicate that the generalizations to 5-D, characterized by Eqs. \ref{eq:operator5D}, \ref{eq:commutation5D}, \ref{eq:pi_timetempaxis}, \ref{eq:hamil_bothaxis},
are backward compatible with the existing 4-D imaginary time formalism.
A fairly generic time-dependent Hamiltonian density can be written as:
\begin{equation}
        H_I(t) = \int \sH_I(\bx,t)d^3x,
\end{equation}
where,
\begin{equation}
        \label{eq:timedephamiltoniandensity}
        \sH_I(\bx,t) = \sum_i c_i(\bx,t)f_i(\hat{\phi},\partial_{\mu}\hat{\phi}, \hat{\pi},\partial_{\mu}\hat{\pi}),
\end{equation}
and $c_i(\bx,t)$ are arbitrary scalar functions.
However, variations in $c_i(\bx,t)$, should not be sharp enough to throw the system out of local thermal equilibrium.
In Ref.~\cite{gans7}, the thermodynamic properties were evaluated for the below specific cases:
\begin{enumerate}
        \item $\sH_I(\bx,t) = V(\bx,t) \hat{\phi}^2$,
        \item $\sH_I(\bx,t) = \lambda(\bx,t) \hat{\phi}^4$.
\end{enumerate}
The evolution operator $U_{\beta}(H_{\beta}, \beta,t)$ provides the evolution w.r.t. $\beta$, i.e.,
\begin{equation}
        U_{\beta}(H_{\beta},\beta,t) = \exp \left ( - \int \int_0^{\beta(\bx,t)} \sH_{\beta}(\bx,t) d\tau d^3x \right ).
\end{equation}
As mentioned earlier, we drop the subscript $\beta$ from $U$, $H$ and  $\sH$, and obtain,
\begin{multline}
        \label{eq:betaevolve}
        U(H,\beta,t) = \exp \left (- \int \int_0^{\beta(\bx,t)}  \sH(\bx,t) d\tau d^3x \right ),\\
        =  \exp \left (- \int \int_0^{\beta_0} s(\bx,t) \sH(\bx,t) d\tau d^3x \right ),
\end{multline}
where, $\sH(\bx,t) = \sH_0(\bx) + \sH_I(\bx,t)$.
The integral in Eq.~\ref{eq:betaevolve} is akin to evaluating $\sH(\bx,t)$ in a curved 5-D space, with metric $[-s(\bx,t)^2,1,-1,-1,-1]$ and volume element $     s(\bx,t)d\tau d^3x$, at time slice $t$. 
We can then define the partition function $Z(\beta_0,t)$, at a time slice, $t$, in 5-D space-time-temperature as:
\begin{equation}
        \label{eq:partition}
        Z(\beta_0, t) = tr \big [\langle \phi_f| U(\sH,\beta_0,t)|\phi_0 \rangle \big ],
\end{equation}
with, $|\phi_0\rangle$ and $|\phi_f\rangle$ being the eigenkets at $\tau = 0$ and $\tau=\beta_0$ respectively.

Finally, the partition function after evaluation using path integral methods, becomes~\cite{gans7},
\begin{multline}
\label{eq:pathintpart}
        Z(\beta_0,t) = trK(\phi_f,\phi_0,\beta_0,t) =\\
        \int_{periodic} D\phi \int \frac{D\pi}{2\pi}
        \exp \Bigg \{ \int_0^{\beta_0} d\tau \int d^3x  \sqrt{g_5(\bx,t)}\\
\times  \Big [ i\pi(\bx,\tau)\frac{1}{{s(\bx,t)}}D_{\tau} \phi(\bx,\tau)  \\
        -  \sH(\pi(\bx,\tau),\phi(\bx,\tau),t)  \Big ]  \Bigg \}.
\end{multline}
The above expression is precisely what one would have obtained if one had considered a space with the metric $G^5 = diag[-s(\bx,t)^2,1,-1,-1,-1]$. This is a curved 5-D space-time-temperature.
Further, for the metric, $G^{5} = diag[-s(\bx,t)^2,1,-1,-1,-1]$. We have, $\sqrt{|G^{5}|} = s(\bx,t)$. We denote the determinant of the Euclidean metric, $|G^{5}|$, by $g_5$. 
This indicates that the variation in the temperature can be recast as a variation in the metric.
Additional intuition and justification in terms of Polyakov loop, and AdS-CFT duality, are provided in Ref.~\cite{gans6}.

\subsection{Einstein field equations}
We now touch upon the combined effects of both gravity and temperature variations, first introduced in Ref.~\cite{gans7}.
We use the letters, $a$ and $b$, as indices for the 5-D space-time, i.e., $a$, $b$ = 0, 1, 2, 3, 4, with the index 0 referring to the temperature dimension, and the index 1 referring to the time dimension,
The letters, $\mu$ and $\nu$, are indices for the 4-D Lorentzian space-time, i.e., $\mu$, $\nu$ = 1, 2, 3, 4.
A superscript, $(N)$, within brackets, refers to $N$ dimensional space. For example, $\nabla^{(4)}_{\mu}$, refers to the covariant derivative in 4-D space-time.

Based on Eq.~\ref{eq:8Dmetric},
we consider the 5-D metric, $g^{(5)}_{ab}$, as:
            \begin{equation}
                    \label{eq:5Dmetric_v0}
                    g^{(5)}_{ab} =
                \left[ \begin{array}{c c}
                        s^2(g^{(4)}_{\mu\nu}(\bx,t),\bx,t)& 0\\
                                    0 & g^{(4)}_{\mu\nu}\\
                \end{array} \right ],
            \end{equation}
where, $g^{(4)}_{\mu\nu}$, is the usual metric tensor in 4-D space-time, and is purely due to gravitational fields.
It is also noted that $s$ would now additionally be a function of $g^{(4)}_{\mu\nu}$ also.
For the metric, $g^{(5)}_{ab}$, the Einstein field equations in the 5-D space become:
\begin{equation}
        \label{eq:EFE5D_v0}
        R^{(5)}_{ab} - \frac{1}{2}g^{(5)}_{ab} R^{(5)} = \frac{8\pi G}{c^4} T^{(5)}_{ab},
\end{equation}
where,
\begin{equation}
        T^{(5)}_{ab} = -\frac{2}{\sqrt{g^{(5)}}} \frac{\delta S}{\delta g^{(5)}_{ab}},
\end{equation}
with, $S$, being the action.
Finally the 5-D Ricci scalar, $R^{(5)}$ can be expressed as:
        \begin{equation}
                R^{(5)} = R^{(5)a}_{~~a} = R^{(4)} - \frac{2}{s} \nabla^{(4)\mu}\nabla^{(4)}_{\mu} s.
        \end{equation}

	In Ref.~\cite{gans7}, Eq.~\ref{eq:EFE5D_v0} was shown to predict new phenomena such as the spontaneous symmetry breaking of scalar fields in the presence of a strong gravitational field.

This concludes the overview of the theory developed in Refs.~\cite{gans6,gans7}, and we now apply the theory (Eq.~\ref{eq:EFE5D_v0} in particular) to galactic systems. 
\section{Formulation}
\label{sec:formulation}
	Based on the theory developed in~\cite{gans7}, we consider the 5-D metric, $g^{(5)}_{ab}$, as:
            \begin{equation}
		    \label{eq:5Dmetric}
		    g^{(5)}_{ab} = 
                \left[ \begin{array}{c c}
			s^2(\bx,t)& 0\\
				    0 & g^{(4)}_{\mu\nu}\\
                \end{array} \right ],
            \end{equation}
where, $g^{(4)}_{\mu\nu}$, is the usual metric tensor in 4-D space-time, and is purely due to gravitational fields, while, $s(\bx,t)$ captures the variation in the inverse temperature, $\beta$.
The sign convention used is (+,-,+,+,+).
	For a system that is in thermal equilibrium, the Einstein field equation in 5-D space-time-temperature is given by Eq.~\ref{eq:EFE5D_v0} (Ref.~\cite{gans7}), i.e.: 
\begin{equation}
	\label{eq:EFE5D}
	R^{(5)}_{ab} - \frac{1}{2}g^{(5)}_{ab} R^{(5)} = \frac{8\pi G}{c^4} T^{1(5)}_{ab},
\end{equation}
where, the stress energy tensor, $T^{1(5)}_{ab}$, is given by:
\begin{equation}
	T^{1(5)}_{ab} = (\rho + \frac{P_1}{c^2})u_a u_b + P_1g^{(5)}_{ab},
\end{equation}
	with, $\rho$ being the density, and $P_1$, the pressure.
The Ricci tensor, $R^{(5)}_{ab}$, can be expressed in terms of the 4-D covariant derivative operator, $\nabla^{(4)}_{\mu}$, and the 4-D Ricci tensor, $R^{(4)}_{\mu\nu}$, as:
            \begin{equation}
		    \label{eq:ricci5D}
		    R^{(5)}_{ab} = 
                \left[ \begin{array}{c c}
			R_{\beta\beta} & 0\\
				    0 & R^{(4)}_{\mu\nu} - \frac{1}{s}\nabla^{(4)}_{\mu} \nabla^{(4)}_{\nu} s\\
                \end{array} \right ],
            \end{equation}

where, 
	$ R_{\beta\beta} = -s\nabla^{(4)\mu} \nabla^{(4)}_{\mu} s $.
	In contrast, for systems that are non-interacting and in complete non-equilibrium, the particles behave as if there are no other particles. 
Without an ensemble or temperature concept, the temperature dimension becomes meaningless. 
To model this scenario, let us take thermal gradient $\rightarrow 0$, i.e., $\partial_{\mu}s \rightarrow 0$, followed by $s\rightarrow 0$. 
A system with zero temperature cannot have thermal gradients. This mandates $\partial_{\mu}s \rightarrow 0$. $\partial_{\mu}s \rightarrow 0$ leads to a system with uniform temperature. Subsequently, $s=0$, eliminates the temperature dimension from the metric $dS^2 = s^2d\beta^2 - dt^2 + dx^2 + dy^2 + dz^2$.
As, $\partial_{\mu}s \rightarrow 0$, it is evident that the 5-D Ricci tensor in Eq.~\ref{eq:ricci5D}, is reduced to a 4-D Ricci tensor. The reduction to 4-D field equations is explained in more detail in Ref.~\cite{gans7}. Moreover, the reduction of 5-D thermal field theory to the standard 4-D imaginary time formalism or the normal 4-D space-time quantum field theory is also covered in Ref.~\cite{gans7}.
One may also refer to Eqs. \ref{eq:pcreation} to \ref{eq:hamiltonian4D}, for the relations between the field theories in 5-D and 4-D.
	After reduction to 4-D, the 4-D Einstein's field equations, represented in 5-D, are:
            \begin{equation}
	\label{eq:EFE4D}
                \left[ \begin{array}{c c}
			0 & 0\\
				    0 & R^{(4)}_{\mu\nu} \\
                \end{array} \right ]
		    -\frac{1}{2} \left[ \begin{array}{c c}
			0 & 0\\
				    0 & g_{\mu\nu} R^{(4)} \\
                \end{array} \right ]
   		    = \frac{8\pi G}{c^4} \left[ \begin{array}{c c}
			0 & 0\\
				    0 & T^2_{\mu\nu} \\
                \end{array} \right ].
            \end{equation}
		    The stress energy tensors, $T^1$ and $T^2$, have the same $\rho$, but different pressures, $P_1$ and $P_2$. $P_1$ is the pressure due to an ensemble interacting gravitation-ally, while $P_2 \sim 0$ in the absence of any ensemble (a single particle has no concept of pressure).

	                An equation representing the behavior of a partially thermalized system, needs to be a generalization of Eqs.~\ref{eq:EFE5D} and~\ref{eq:EFE4D}, with  Eqs.~\ref{eq:EFE5D} and~\ref{eq:EFE4D} being special cases. 
			To motivate the generalization, let us rewrite Eq.~\ref{eq:EFE5D} by taking all terms that depend on $s$ to the R.H.S.. This is inspired by the fact that $s$ is related to inverse temperature, and thus related to inverse energy, and consequently, can be interpreted to act as a source causing space-time curvature.
		    We also apply the simplification that, for a time-invariant system, $u^0 = c\frac{\partial \beta}{\partial \tau_p} = 0$, where $\beta$ is the inverse temperature, and $\tau_p$ is the proper time. Equation~\ref{eq:EFE5D} then becomes,
\begin{multline}
	\label{eq:EFE5D_2}
                \left[ \begin{array}{c c}
			0 & 0\\
				    0 & R^{(4)}_{\mu\nu} \\
                \end{array} \right ]
		    -\frac{1}{2} \left[ \begin{array}{c c}
			0 & 0\\
				    0 & g_{\mu\nu} R^{(4)} \\
                \end{array} \right ] = \frac{8\pi G}{c^4} \\
		\times \left[ \begin{array}{c c}
			\frac{s^2c^4R^{(4)}}{16\pi G} + s^2P_1 & 0\\
				    0 & T^1_{\mu\nu} + \frac{c^4\left (\frac{1}{s}\nabla_{\mu} \nabla_{\nu} s - \frac{g_{\mu\nu}}{s}\nabla^{\alpha} \nabla_{\alpha} s \right )}{8\pi G} \\
                \end{array} \right ],
\end{multline}
where, $\alpha = 1,2,3,4$, and we have skipped the superscript $(4)$ in $\nabla^{(4)}$ for simplicity of notation. 
The terms in the R.H.S. of Eq.~\ref{eq:EFE5D_2}, can be viewed as new source terms describing a thermalized system, with thermal gradients, which causes curvature of 4-D space-time.
We now hypothesize that the source term of a partially thermalized system may be considered to be a linear combination of the source terms in Eqs.~\ref{eq:EFE4D} and ~\ref{eq:EFE5D_2}. In other words,
\begin{multline}
	\label{eq:lincomb}
	T^{p(5)}_{ab} = 
		 (1-k) \left[ \begin{array}{c c}
			0 & 0\\
				    0 & T^2_{\mu\nu} \\
                \end{array} \right ] + k \\
		     \times \left[ \begin{array}{c c}
			     \frac{s^2c^4R^{(4)}}{16\pi G} + s^2P_1 & 0\\
				    0 & T^1_{\mu\nu} + \frac{c^4 \left (\frac{1}{s}\nabla_{\mu} \nabla_{\nu} s - \frac{g_{\mu\nu}}{s}\nabla^{\alpha} \nabla_{\alpha} s \right ) }{8\pi G} \\
                \end{array} \right ].
\end{multline}
		    Here, $k$ represents the degree or extent of equilibration. 
		    Additional interpretation on $k$ in the context of a galaxy, is provided in Sec.~\ref{sec:interpret_k}.
The corresponding 5-D Einstein field equations for a partially thermalized system are then,
\begin{equation}
	\label{eq:EFE5D_p}
                \left[ \begin{array}{c c}
			0 & 0\\
				    0 & R^{(4)}_{\mu\nu} \\
                \end{array} \right ]
		    -\frac{1}{2} \left[ \begin{array}{c c}
			0 & 0\\
				    0 & g_{\mu\nu} R^{(4)} \\
		    \end{array} \right ] 
		    = \frac{8\pi G}{c^4} T^{p(5)}_{ab}.
\end{equation}
If $k$ is very small (as in a galaxy with hindsight), then, $1-k \sim 1$, and Eq.~\ref{eq:EFE5D_p} can be written as
\begin{equation}
	\label{eq:gal1a}
	-\frac{1}{2} k s^2 R^{(4)} = \frac{8\pi G}{c^4}kP_1s^2,\\
\end{equation}
and,
\begin{equation}
	\label{eq:gal1b}
	R_{\mu\nu} - \frac{k}{s}\nabla_{\mu}\nabla_{\nu}s - \frac{1}{2}g_{\mu\nu}R^{(4)} + \frac{kg_{\mu\nu}}{s} \nabla^{\alpha}\nabla_{\alpha}s = \frac{8\pi G}{c^4}T_{\mu\nu},
\end{equation}
where, 
$T_{\mu\nu} =  (\rho + \frac{P_2}{c^2} + \frac{kP_1}{c^2})u_{\mu}u_{\nu} + g_{\mu\nu} (P_2   + kP_1)$.
The form of the pressure terms in $T_{\mu\nu}$, suggests, one can define an effective pressure, 
\begin{equation}
	P_{eff} = P_2 + kP_1. 
\end{equation}
	For the sake of completeness, we note that, if $k$ were not small, one would have obtained: 
\begin{equation}
	\label{eq:Peff}
	P_{eff} = (1-k)P_2 + kP_1.
\end{equation}
$P_{eff}$ can be interpreted as the pressure of a partially  thermalized system, with degree of equilibration, $k$.
If $k$ is small one can neglect $kP_1$, and given that $P_2 \sim 0$,
Eq.~\ref{eq:gal1b} reduces to,
\begin{equation}
	\label{eq:gal3}
	R_{\mu\nu} - \frac{k}{s}\nabla_{\mu}\nabla_{\nu}s - \frac{1}{2}g_{\mu\nu}R^{(4)} + \frac{kg_{\mu\nu}}{s}\nabla^{\alpha}\nabla_{\alpha}s = \frac{8\pi G}{c^4} \rho u_{\mu} u_{\nu}.
\end{equation}
Making use of the fact that $u_1 = c\gamma$, the time component ($\mu=\nu=1$) of Eq.~\ref{eq:gal3}, becomes:
\begin{equation}
	\label{eq:gal4}
	R_{11} - \frac{k}{s}\nabla_{1}\nabla_{1}s -  \frac{1}{2} g_{11}R^{(4)} + \frac{kg_{11}}{s}\nabla^{\alpha}\nabla_{\alpha} s = \frac{8\pi G}{c^2} \rho \gamma^2.
\end{equation}
Since, the velocity of stars is very small, We take $\gamma \approx 1$.

In the weak field approximation, $g_{\mu\nu} = \eta_{\mu\nu} + h_{\mu\nu}$, with $||h_{\mu\nu}||\ll 1$. If additionally, $h_{\mu\nu}$ is time-invariant, i.e., the time derivatives are 0, we then have,
\begin{eqnarray}
	\label{eq:gal5}
R_{11} + \frac{1}{2}R^{(4)} \approx -\frac{1}{2} \nabla^2 h_{11}  + \frac{1}{2} \left ( -\nabla^2 h + \partial_{\alpha} \partial_{\beta} h^{\alpha\beta} \right ),
\end{eqnarray}
where, $\nabla^2$ represents the Laplacian operator, $\partial_x^2 + \partial_y^2 + \partial_z^2$.
If $k$ is very small, then we can take a perturbation, i.e., $h_{\mu\nu} = -\frac{2\phi_0(\bx)}{c^2} \delta_{\mu\nu} + \Delta h_{\mu\nu}$,
where,
            \begin{equation}
		    \label{eq:5Dmetric_perturb}
		   \Delta h_{\mu\nu} = 
                \left[ \begin{array}{c c}
			\Delta h_{11}& 0\\
				    0 & \Delta h_{ij}\\
                \end{array} \right ],
            \end{equation}
	    with $i,j=2,3,4$.
	                We have ended up introducing an extra variable $\phi_0$.
            In order to unambiguously specify the $h_{\mu\nu}$ split between $-\frac{2\phi_0(\bx)}{c^2} \delta_{\mu\nu}$ and $\Delta h_{\mu\nu}$,
we apply the constraint:
\begin{multline}
\label{eq:gauge}
	\nabla^2 \Delta h_s - \partial_x^2\Delta h_{22} - \partial_y^2 \Delta h_{33} - \partial_z^2 \Delta h_{44} \\
	-2\big ( \frac{\partial^2 \Delta h_{12}} {\partial x \partial y} + \frac{\partial^2 \Delta h_{23}}{\partial y \partial z} + \frac{\partial^2 \Delta h_{13}}{\partial x \partial z} \big ) = 2\Delta h_{11},
\end{multline}
where, $\Delta h_s = \Delta h_{22} + \Delta h_{33} + \Delta h_{44}$.
Then,
\begin{equation}
	\label{eq:gal6}
	R_{11} + \frac{1}{2}R^{(4)} = \frac{2}{c^2} \nabla^2 \phi(\bx),
\end{equation}
where $\frac{2\phi}{c^2} = h_{11} = \frac{2\phi_0}{c^2} - \Delta h_{11}$.
Substituting Eq.~\ref{eq:gal6} in Eq.~\ref{eq:gal4}, and recognizing, $\nabla_{1}\nabla_{1}s \approx \frac{\partial^2s}{\partial t^2} = 0$ for a time-invariant $s$, and $g_{11} \approx -1$, we obtain in the weak field case, 
\begin{equation}
	\label{eq:gal8}
	\nabla^2 \phi(\bx) =   4\pi G \rho(\bx) + kc^2\frac{1}{2s(\bx)}\nabla^2s(\bx).
\end{equation}
At large distances, where $\rho$ becomes very small, the potential is dominated by  $kc^2\frac{1}{2s}\nabla^2s$ term.
For $k=0$, one recovers the Newtonian gravity. 

	A two-body system, where there is no concept of an ensemble, illustrates a system that is quite precisely described by $k=0$.
	The solar system is almost a two-body system, with $k$ being very near zero or zero. The $4\pi G\rho$ term is significant for a solar system, and hence dominates.
	A system with $k=1$, constitutes a thermal system in a gravitational field and, we may call this theory, the theory of thermal gravity. This is essentially, the Einstein field equations in 5-D space-time-temperature as described in~\cite{gans7}.
	A system with $0 < k < 1$, constitutes a partially thermalized, many-body system in a gravitational field. We may call this theory, the theory of many body gravity (MBG). 
The value of $k$ is a free parameter, and is determined by a fit to the data in this paper. 
	One may extend the above scenario, to systems larger than the galaxy. On a much larger cosmic scale, it may be possible to envisage galaxies, as point particles interacting with each other. 
	It is now shown that the MBG theory is able to explain the rotation curves of the Milky Way and the M31 galaxies.

	For the rotation curves, it is useful to determine $\nabla \phi$ in the radial direction.
Once, Eq.~\ref{eq:gal8} is solved, then, the radial component of $\nabla \phi$ is:
\begin{eqnarray}
		\frac{\partial \phi}{\partial r} = \frac{\partial \phi}{\partial x} \cos(\theta) + \frac{\partial \phi}{\partial y} \sin(\theta),
\end{eqnarray}
where, $\theta$ is the angle subtended by the radial vector $\vec{r}$ on the x-axis.

\subsection{The Geodesic Equations}
We solve the geodesic equation for the metric in cylindrical co-ordinates, i.e., for the metric: 
\begin{multline}
	g^{(5)}_{ab} = diag[s^2, -(1+ \frac{2\phi}{c^2}) + \Delta h_{11},  1- \frac{2\phi}{c^2} + \Delta h_{rr},\\
	r^2(1- \frac{2\phi}{c^2} + \Delta h_{\theta\theta}), 1- \frac{2\phi}{c^2} + \Delta h_{zz}] \\
	+ spatial~non-diagonal~elements.
\end{multline}
For a time-invariant galactic system, we assume the condition, $\frac{\partial s}{\partial t} = 0$. Further, if a star has only orbital velocity, then, $\frac{\partial r}{\partial t} = 0$,$\frac{\partial z}{\partial t} = 0$.
Also, close to the galactic plane, we assume that derivatives of the metric components w.r.t $z \sim 0$, due to symmetry above and below the $r-\theta$ plane.

Consequently, for small $\phi_0$ and $\Delta h_{\mu\nu}$, taking the lowest power terms in $\phi_0$ and $\Delta h_{\mu\nu}$, the geodesic equations lead to:
\begin{equation}
	\label{eq:geod1}
	r\left ( \frac{\partial \theta}{\partial t}\right )^2  = \left ( \frac{\partial \phi_0}{\partial r} - \frac{c^2}{2}\frac{\partial \Delta h_{11}}{\partial r}\right ) =  \frac{\partial \phi}{\partial r}.~~~~~~
\end{equation}
Finally,
\begin{equation}
	\label{eq:vel}
	        v^2 = r^2\left ( \frac{\partial \theta}{\partial t} \right )^2 = r\frac{\partial \phi}{\partial r},
\end{equation}
where, $v$ is the rotational speed of the star.

\subsection{Modeling of $s$}
For a partially thermalized system like a galaxy, the stress energy tensor is modeled as a linear combination of the stress energy tensors of a thermalized system in local thermal equilibrium, with variation in inverse temperature, $s$,  and a system in complete non-equilibrium (Ref. Eq.~\ref{eq:lincomb}), 
Though, $s$ appears in Eq.~\ref{eq:gal8}, only the properties of a galaxy can be measured. 
Thus, it is required to relate $s$ to a state variable of a galaxy, which can then be used to determine $s$, or used in place of $s$. 

For a fully thermalized system, the energy expectation value is given by
\begin{equation}
	\langle E\rangle = \frac{1}{Z}\sum_{i}n_i E_i \exp\left (-\frac{E_i}{\sK T}\right ),
\end{equation}
where, $n_i$ particles have energy $E_i$, $Z$ is the partition function, and $\sK$ is the Boltzmann constant.
For a given distribution, $n_i$, it is a bijective mapping between $\langle E\rangle$ and temperature, $T$.
Thus, the temperature of a thermal system is a measure of the energy expectation value of the system. As long as the distribution is constant, one may use the two quantities interchangeably.  
In general, the inverse temperature or $s$, can be generalized to some function of the variation of the inverse energy $\langle E\rangle$, i.e., $s = \frac{1}{f(\langle E\rangle)}$.
The function itself may depend on the distribution of particles, such as Boson gas, Fermion gas, classical system, etc.
Let $\frac{1}{\sK T(\bx)} = \beta(\bx) = s(\bx)\beta_0$, with $\beta_0$ = a constant, be the inverse temperature at a point $\bx$ in the thermalized system (Ref.~\cite{gans6,gans7} for the relation $\beta(\bx) = s(\bx)\beta_0$).
For a classical system in local thermal equilibrium, the expected energy, $\langle E_{cl}(\bx)\rangle$, of a particle at a point $\bx$, can be taken as: $\langle E_{cl}(\bx)\rangle \approx \sK T(\bx)$, 
and one obtains, 
\begin{equation}
\label{eq:sE}
	s(\bx) \approx \frac{1}{\beta_0\langle E_{cl}(\bx) \rangle} .
\end{equation}

Thus, for a thermalized system (the degree of equilibration $k=1$), the energy expectation value and the temperature can be used interchangeably, as they represent the same physics.

For a partially thermalized system ($0 < k < 1$), from Eq.~\ref{eq:Peff}, we have, $P_{eff} = (1-k)P_2 + kP_1$. 
Here, $P_1$ is the pressure of the thermalized system, with inverse temperature, $\beta(\bx) = s(\bx)\beta_0$.
Then, the energy expectation value, $\langle E(\bx)\rangle$, of a particle at a point $\bx$, in a partially thermalized system is,
\begin{multline}
	\label{eq:Ea}
	\langle E(\bx) \rangle = \frac{1}{\Delta n(\bx)}\int_{\Delta V_{\bx}} P_{eff}dV\\
	= \frac{1}{\Delta n(\bx)}\int_{\Delta V_{\bx}} \left \{ (1-k)P_2 + kP_1 \right \} dV,
\end{multline}
where, $\Delta n(\bx)$ is the number of particles within a small volume, $\Delta V_{\bx}$, around the point $\bx$.
In  Sec.~\ref{sec:formulation}, we had taken $P_2 \sim 0$, for a non-interacting system in complete non-equilibrium, like an isolated single particle, or collection of isolated non-interacting particles, as there is no concept of pressure for such a system. We then obtain,
\begin{equation}
	\label{eq:Eb}
	\langle E(\bx) \rangle = \frac{1}{\Delta n(\bx)}\int_{\Delta V_{\bx}} kP_1 dV = k\langle E_1(\bx) \rangle,
\end{equation}
where, $\langle E_1(\bx) \rangle = \frac{1}{\Delta n(\bx)}\int_{\Delta V_{\bx}} P_1 dV$, is the energy of a particle within the small volume, $\Delta V_{\bx}$, existing in a thermalized system at local thermal equilibrium, equilibrated via gravitational interactions. 
If the thermalized system is a classical system in local thermal equilibrium, then from Eq.~\ref{eq:sE}, $\langle E_1(\bx) \rangle \approx  \frac{1}{\beta_0 s(\bx)}$. Finally, from Eq.~\ref{eq:Eb}, we have:
\begin{equation}
	\label{eq:Ec}
	\langle E(\bx) \rangle \approx \frac{k}{\beta_0 s(\bx)}.
\end{equation}

In a gravitational system, like a galaxy, the potential energy of a particle (star) is proportional to the gravitational potential, $\phi$, it sees due to the presence of other particles (stars) in the system. Subsequently, we take the energy expectation value, $\langle E(\bx) \rangle \propto \phi(\bx)$, where $\phi$ is the gravitational potential. 
Putting everything together, we get, $s\propto \frac{1}{\phi}$.
It is to be noted that, since $\frac{1}{s}\nabla^2s$ is scale-invariant, any constant of proportionality in $s$ is irrelevant.
In this case, Eq.~\ref{eq:gal8} becomes:
\begin{equation}
	\label{eq:gal9}
	\nabla^2 \phi -  kc^2\frac{\phi}{2}\nabla^2\frac{1}{\phi} =   4\pi G \rho.
\end{equation}

\subsection{Interpretation of $k$ for a galaxy} 
\label{sec:interpret_k}
Let us touch upon the intuitive reasons why the many-body system may be expected to affect the dynamics of stars in a galaxy. If there is no interaction between the stars, they simply act as individual particles and follow Einstein's general relativity. But in the event of interactions, the stars are not able to act individually. They act like a bunch. The interstellar gasses may augment the grouping effect. Their velocities are affected by the position and velocities of the other stars and gasses within the group. The extent to which they act like a group is determined by the value of $k$.

To elaborate a little bit more, consider a particle moving under the influence of Newtonian gravity. It would have a deterministic velocity $v(t)_{deter}$. A particle which is part of a fully thermalized system would have a random velocity expectation value $\langle v(t)_{rand} \rangle$. But a particle which has a combination of random velocity and deterministic velocity could have the final velocity as:
\begin{equation}  
	\label{eq:linvel}
	v_{final} = (1-k)v(t)_{deter} + k\langle v(t)_{rand} \rangle,
\end{equation}  
where, $k$ can be a constant, and determines the degree of mixing.
However, if instead of velocities, the energies are linearly added, then the relation would be:
\begin{equation}  
	\label{eq:linenergy}
	v^2_{final} = (1-k)v(t)^2_{deter} + k\langle v(t)^2_{rand} \rangle.
\end{equation}  
In the case of the stars in a galaxy, the $k\langle v(t)_{rand}^2\rangle$ obtains a non-zero value due to the random interaction with the other stars and gasses.
Instead of adding energies which are scalars, Eq.~\ref{eq:lincomb} incorporates a linear combination of the tensor form of the energy, namely, the stress energy tensors. 
The ramifications of $v_{rand}$ and $v_{deter}$ can be different.
As an example, a bunch of particles in random thermal motion can have a tendency to fly away from each other (a negative Ricci scalar). On the other hand a bunch of particles under each other's gravitational pull, will have a radial $v_{deter}$ towards each other (a positive Ricci scalar). A real system may be a combination of both the effects. 
An explosive with an energy of explosion, E, would expand. But, the gravitational effect of energy $E$, would try to contract, although a much weaker effect. The resultant effects of these two phenomena is captured by the MBG theory, with $0<k<1$. $k=1$ is a pure explosion, while $k=0$ is pure Einstein gravity. $0<k<1$ indicates a combination of the two phenomena. 
An implosive process on the other hand would have a positive Ricci scalar, and reinforce the gravitational force.
In general, the random thermal part can lead to a positive or negative contribution to the Ricci scalar based on the sign of the second derivative term, i.e., the $\frac{1}{s}\nabla^2 s$ term in Eq.~\ref{eq:gal8}.

The explosion or implosion is an extreme case. Let us instead moderate the random velocity effects significantly, by reducing $k$.
Let us also say that the reduction is to such an extent, that the energy of the random component is comparable or lesser than the gravitational energy.
In this case, a particle rotating around the gravitational center of mass, would see a decreased or increased pull towards the center, leading to a decreased or increased rotational velocity.

In a galaxy, stars acquire a non-zero random velocity due to interaction with other stars and gasses. The effect of such interactions is captured by a non-zero $k$ in the MBG. This is however, a very feeble effect, and hence is manifest only when gravity is very weak.

Another point of view would be to look at the equivalence between the thermal and gravitational accelerations.
Let us say that there is an ant, confined to a safe and sealed, sound proof, and thermally insulated titanium canister placed inside the explosive material. The ant has no way to know whether it is accelerating due to an explosion or due to gravity, essentially suggesting an equivalence between the two accelerations.
Thus, both the phenomena should lead to curvature of space-time. 

A completely different point of view would be to consider a thermalized state as a state of maximal entropy. Equation~\ref{eq:lincomb} then depicts a state, that is a linear combination of energy and some function of entropy. 
An analogy can be found in thermodynamics, where one defines the free energy as a linear combination of energy and entropy, i.e., 
\begin{equation}
\label{eq:free_energy}
\Delta G_0 = \Delta E - \Theta \Delta \sS',
\end{equation}
where, $G_0$, $E$, $\Theta$, $\sS'$, are the Gibbs free energy, energy, temperature and entropy respectively.

The inference of Eq.~\ref{eq:gal9} is that the whole is more than the sum of the individual particles or stars.
If the Newtonian gravity is extremely feeble, then the group effect can dominate. 

\subsection{Tully Fisher relation}
\label{sec:tfr}
At large distances in a galaxy, the mass density, $\rho$, is negligible. For cases where $\rho$ can be neglected, Eq.~\ref{eq:gal9} can be written as:
\begin{equation}
	\label{eq:gal10}
	\nabla^2 \phi \approx  kc^2\frac{\phi}{2}\nabla^2\frac{1}{\phi}.
\end{equation}

The term $\phi \nabla^2\frac{1}{\phi}$ is scale invariant, i.e., if we scale $\phi$, the term $\phi\nabla^2\frac{1}{\phi}$ does not scale.    
Consequently, based on Eq.~\ref{eq:gal10}, in terms of scaling, $\nabla^2\phi \propto k \Rightarrow \nabla \phi \propto k$.
Near the galactic plane, assuming symmetry above and below the galactic plane, $\frac{\partial \phi}{\partial z} \approx 0$.  Then, in cylindrical co-ordinates, $\nabla \phi \approx \frac{\partial \phi}{\partial r}$.
From Eq.~\ref{eq:vel}, we then obtain,
\begin{equation}
\label{eq:gal12}
v^2 = r\frac{\partial \phi}{\partial r} \propto k.
\end{equation}
The interaction a star experiences with the rest of the galaxy would depend on the total number of stars in the galaxy and their masses.  
Therefore, the degree of interaction, $k$, is expected to be a function of the galactic mass, $M$.
If $k \propto \sqrt{M}$, we reproduce the Tully Fisher relation~\cite{tfr}. 
Thus, empirically, we premise:  
\begin{equation}
	\label{eq:tfr}
k \propto \sqrt{M}.
\end{equation}
	In fact, in Sec.~\ref{sec:rar}, it is shown that $k \propto \sqrt{M}$ is able to capture the trend spanning three orders of magnitude in galactic masses.
	In appendix~\ref{sec:tully_appendix}, mathematical arguments are presented, which show that the relation between $k$ and $M$, should resemble Eq.~\ref{eq:tfr}.
	However, the relation $k \propto \sqrt{M}$ requires deeper understanding.



\section{Numerical Simulations and Results for Milkyway and M31}
\label{sec:numerical}
		Equation~\ref{eq:gal9} is a non-linear partial differential equation (PDE) in $\phi$. To make the solution of Eq.~\ref{eq:gal9} more tractable, and allow numerical techniques like the discrete cosine transform (DCT), we apply an initial estimate, $s= \frac{1}{\phi_{clas}}$. Here, $\phi_{clas}$ is obtained by solving the classical Einstein field equations, i.e., $\nabla^2\phi_{clas} = 4 \pi G\rho$. 
		The potential, $\phi$, obtained by solving Eq.~\ref{eq:gal8}, with $s= \frac{1}{\phi_{clas}}$, then becomes a first order modification of $\phi_{clas}$.
		We then iterate over the obtained function, $\phi$, i.e., $s$ is assigned the newly obtained value of $\phi$, and subsequently, Eq.~\ref{eq:gal8} is again solved. In this work, the results after four of these iterations, have provided a reasonable fit to the observed galaxy rotation curve, 

	We solved Eq.~\ref{eq:gal8}, with $s=\frac{1}{\phi}$, using 3D DCT with Octave (a clone of Matlab). DCT was preferred over Fast Fourier transform (FFT), since DCT returns real-valued coefficients. 
A grid size of $256\times256\times256$ was used at the start, with a sampling rate of 0.5kpc per sample. 
In each successive iteration, we reduce the grid size, in order to remove errors at the boundary of the grid, probably due to finite size grid effects.
		The grid size for the fourth iteration was $150\times150\times 150$.
The galactic mass distribution was modeled as a 2D disc in the $z=0$ plane. We model the mass distribution in only one octant. 
The 3D DCT, by its very nature, symmetrizes the distribution by mirroring the values in one octant onto the other octants. Thus, by default, a symmetrical 3D distribution is obtained, even though the mass distribution is explicitly modeled in a single octant. This approach enhances the quality of the numerical results.
In the first iteration, we take $s = \frac{1}{\phi_{clas}}$ in Eq.~\ref{eq:gal8}, and obtain $\phi_1$.
In the next iteration, assign $s = \frac{1}{\phi_1}$, and obtain $\phi_2$, and so on. 
To facilitate the use of DCT, we have chosen $(x,y,z)$ co-ordinate system, even though, $\rho$ has rotational symmetry around the z-axis. This brings about some numerical issues. In order to minimize the impact of such numerical errors, we take the value of $\phi$, and $\nabla \phi$, along the diagonal line, $x=y$, in the plane, $z=0$.
We now present the results for the Milky way and the M31 galaxies.
\\
\begin{center}
\begin{table}
\begin{tabular}{ |c|c|c|c|c|c|c|c|c|  }
 \hline
	\multirow{2}{4em}{Galaxy} & \multicolumn{4}{|c|}{Bulge } & \multicolumn{3}{|c|}{Disc }& BH \\
	\cline{2-9}
	& $M_b$     & $\rho_{0b}$                  & n&$R_b$ & $M_d$      & $\rho_{0d}$    &$R_d$& $M_{BH}$\\
	& 1e10      & 1e11                       &  & (kpc)& 1e10       & 1e6         &(kpc) & 1e6 \\
	&$M_{\odot}$& $\frac{M_{\odot}}{kpc^{2}}$&  &      & $M_{\odot}$& $\frac{M_{\odot}}{kpc^{2}}$& &$M_{\odot}$\\
 \hline
	Milkyway& 1.8* & 68*    & 4*  & 0.5*&6.5* & 844* &3.5*& 3.7* \\
 \hline
	M31     & 3.1$^{\dagger}$ & 1.063 & 2.2**& 1.0**&5.6$^{\dagger}$ & 324 & 5.3**&80$^{\dagger}$ \\
	\hline
\end{tabular}
	\caption{{*} Reference~\cite{sofue2}; {**} Reference~\cite{andro1}; {${\dagger}$ }  Reference~\cite{andro2}.} 
\label{tab:parameters}
\end{table}
\end{center}

\begin{figure}
\includegraphics[width = 80mm,height = 80mm]{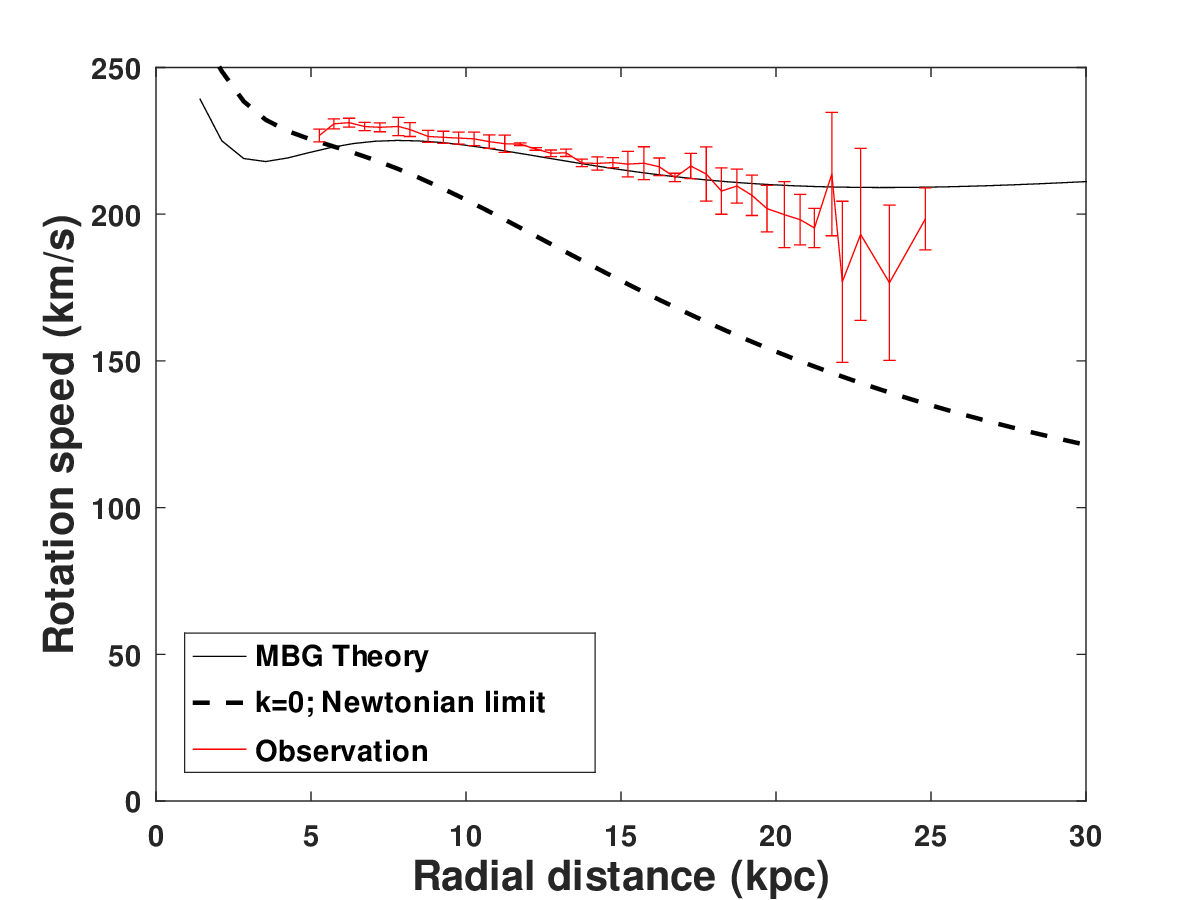}
        \caption{Milky way rotation curves:
	root mean square (r.m.s.) error = $\sigma_{MBG} = 9.3899 km/s$,
	$\sigma_{Newton} = 37.649 km/s$
	} 
\label{fig:milkyway}
\end{figure}
\begin{figure}
\includegraphics[width = 80mm,height = 80mm]{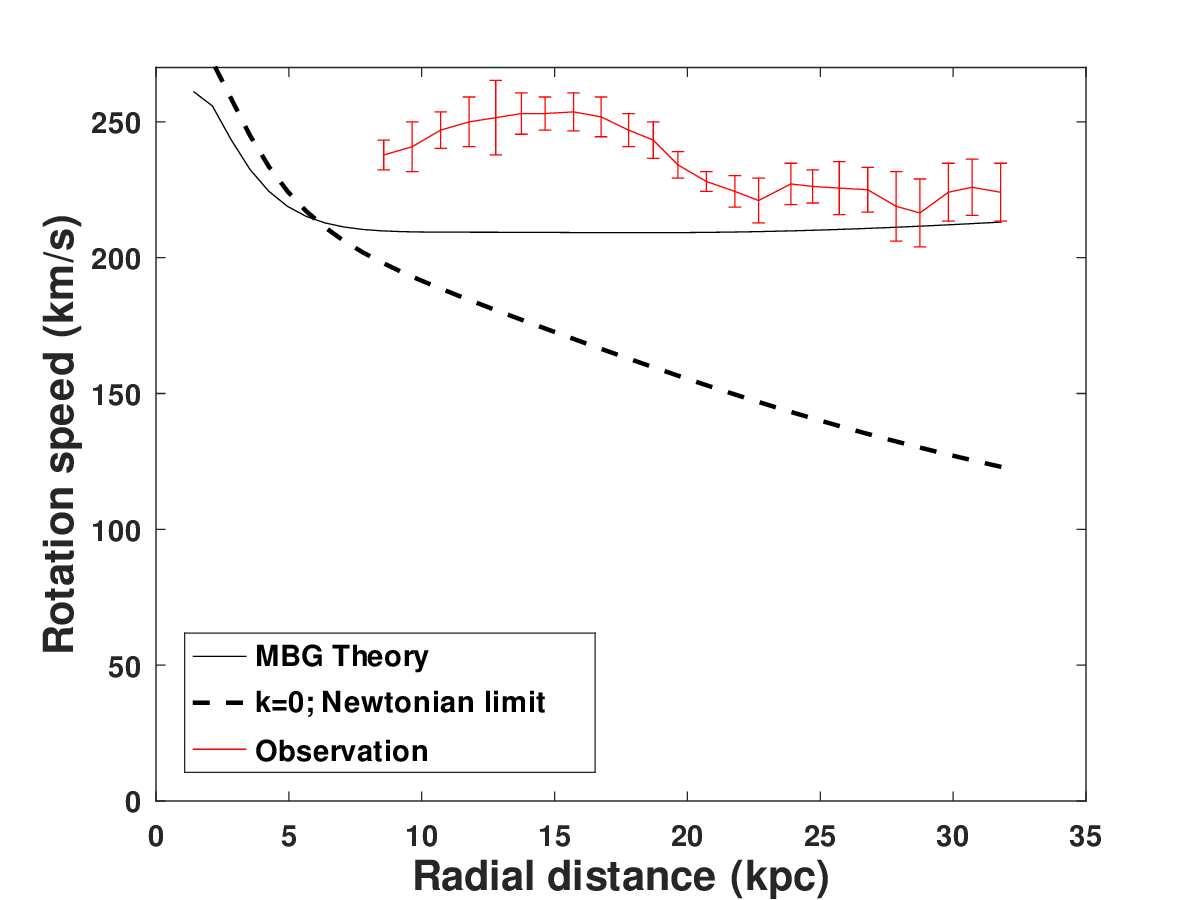}
        \caption{M31 rotation curves:
	r.m.s. error = $\sigma_{MBG} = 27.980 km/s$,
	$\sigma_{Newton} = 83.076 km/s$
	} 
\label{fig:andromeda}
\end{figure}
	The values in the table~\ref{tab:parameters}, have been obtained from Refs.~\cite{sofue2,andro1,andro2}. The parameters, $\rho_{0b}$ and $\rho_{0d}$, have been calculated for M31 based on the other M31 parameters in table~\ref{tab:parameters}, and assuming a flat 2D mass profile. 
	The sersic profile is used to model the mass distribution.
	The mass profile in the disc region is taken as exponential.
	The observed experimental data (red line) for the Milky way in Fig.~\ref{fig:milkyway} is obtained from~\cite{hammer} and the observed data(red error bars) for the M31 in Fig.~\ref{fig:andromeda} is obtained from~\cite{andro2}. 
	Based on Eq.~\ref{eq:tfr}, the value of $k$ in the simulation is, 
$k = 2.8157\times 10^{-11}\sqrt{M_{galaxy}}$, 
where $M_{galaxy} = M_{BH} + M_b + M_d$. 
	Although the value of $k$ is very small, the $\frac{kc^2}{2s}\nabla^2s$ term is able to dominate at large radial distances, as the Newtonian gravity is very feeble.
	Overall, in Figs.~\ref{fig:milkyway} and~\ref{fig:andromeda}, 
	the simulated curves from the proposed MBG theory, lie relatively close to the observed experimental values (red curve) to a reasonable extent, especially at larger radial distances. 
	The $k=0$ curve provides the result based on the usual Newtonian gravity and significantly underestimates the galaxy rotation curve.

\section{NFW curve, RAR, WBS and other phenomenon}
	\label{sec:prediction}
	\subsection{The pseudo mass and the NFW curve}
	Let $\nabla^2\phi = 4\pi G \rho + \frac{kc^2}{2}\phi \nabla^2\frac{1}{\phi} = 4\pi G \rho_{eff}$.
	Then, $\Delta \rho = \rho_{eff} - \rho = \frac{kc^2}{8\pi G}\phi \nabla^2\frac{1}{\phi}$, becomes a pseudo mass, the analogue of dark matter in the proposed MBG theory.
	The value of the pseudo mass would depend on the value of $\nabla^2\frac{1}{\phi}$, or $\square \frac{1}{\phi}$, in the case of time-varying systems, and also on the value of $k$.

\begin{figure}
\includegraphics[width = 80mm,height = 80mm]{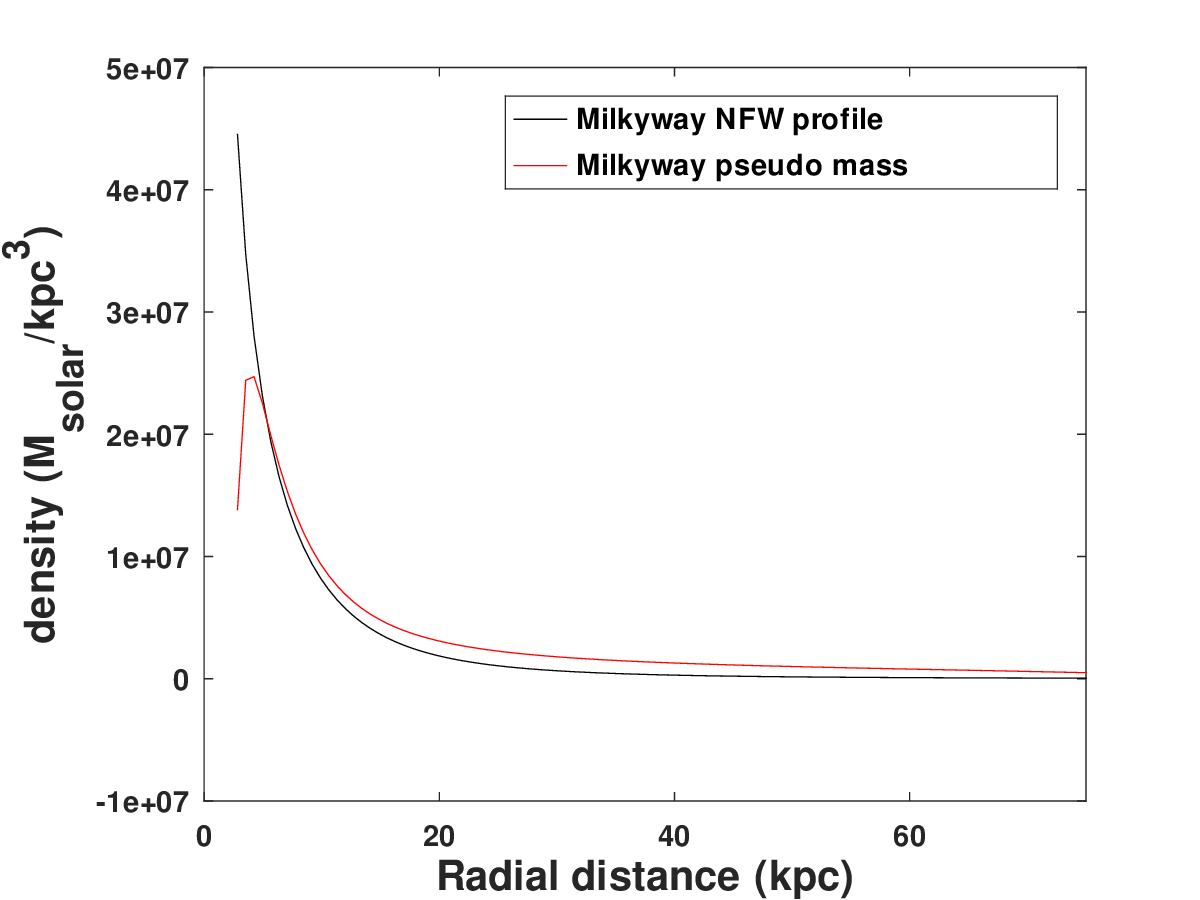}
        \caption{Milkyway: Comparison of the NFW dark matter profile with the pseudo mass curve.
	The pseudo mass curve depicts the density along the galactic plane.
        }
\label{fig:milkyway_nfw}
\end{figure}
\begin{figure}
\includegraphics[width = 80mm,height = 80mm]{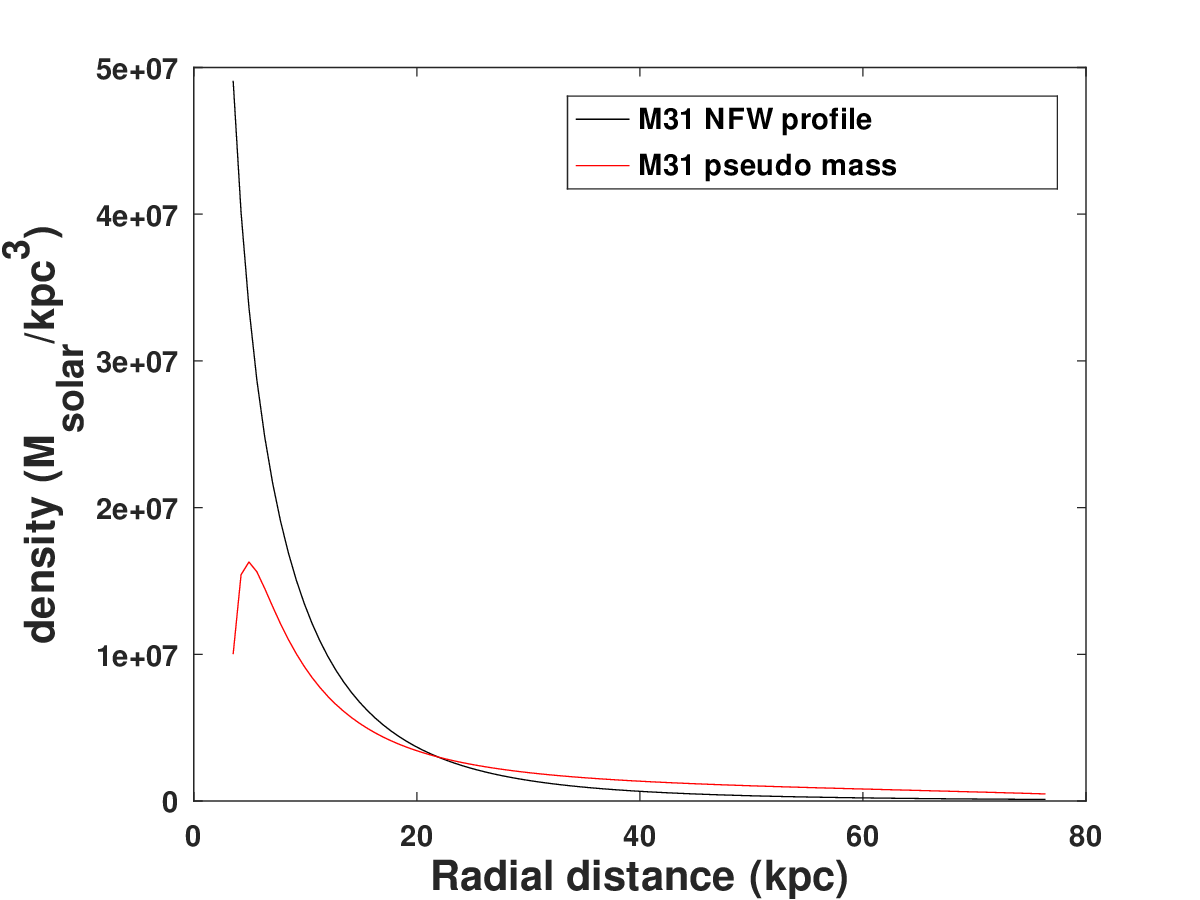}
	\caption{M31: Comparison of the NFW dark matter profile with the pseudo mass curve. 
	The pseudo mass curve depicts the density along the galactic plane.
        }
\label{fig:andromeda_nfw}
\end{figure}
The Navarro–Frenk–White (NFW)~\cite{nfw} profile is a commonly used profile to model the dark matter distribution.
In Fig.~\ref{fig:milkyway_nfw}, the NFW curve for the Milkyway is compared with the Milkyway's pseudo mass profile. The NFW curve for Milkyway was obtained from Ref.~\cite{mw_nfw}. The pseudo mass curve depicts the density along the galactic plane. 
The NFW profile and the pseudo mass curve along the galactic plane for M31 are compared in Fig.~\ref{fig:andromeda_nfw}. The NFW curve is based on the M31 NFW parameters from Ref.~\cite{and_nfw}. 
The NFW and the pseudo mass curves are comparable for both the Milkyway and the M31.
\begin{figure}
\includegraphics[width = 80mm,height = 60mm]{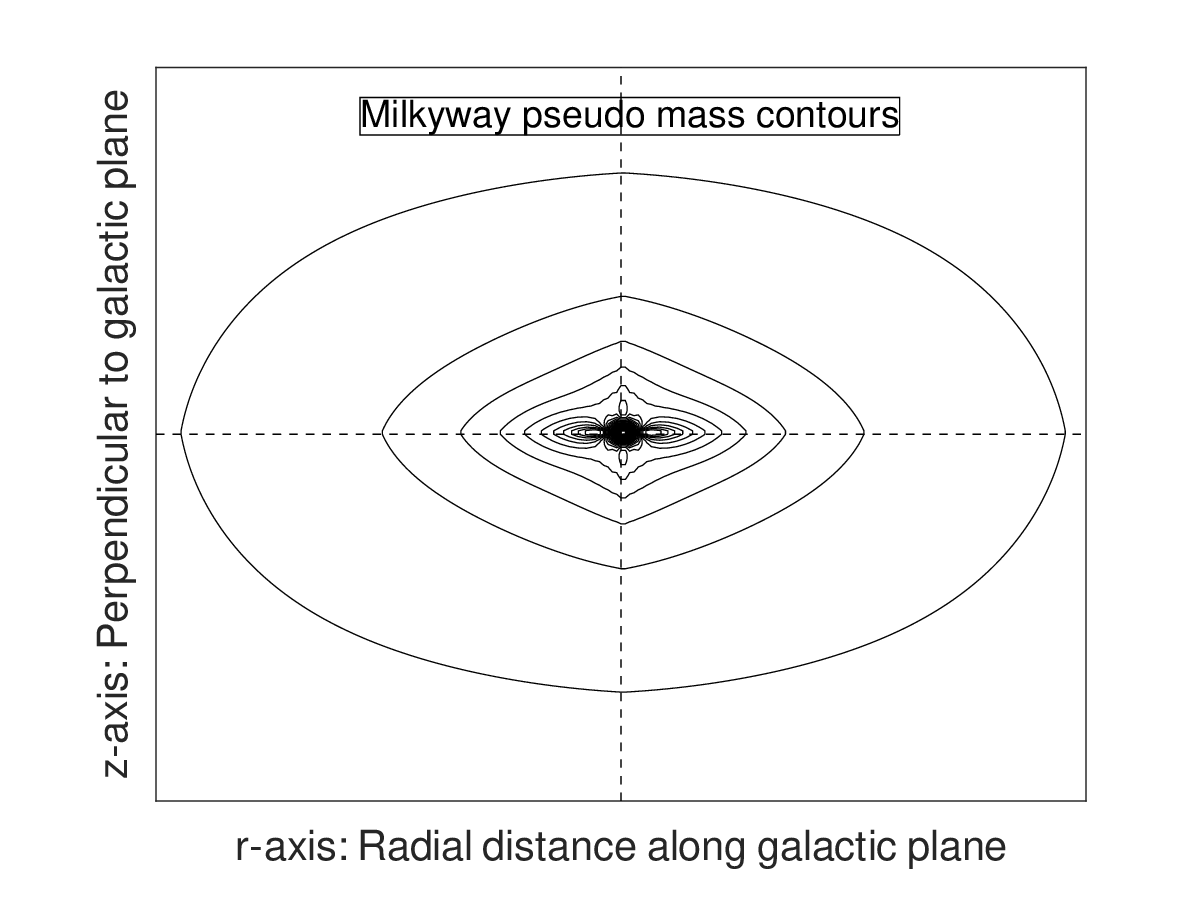}
        \caption{Milkyway: Pseudo mass contours on a plane perpendicular to the galactic plane. 
	At the intersection of the $r$ and $z$-axis (dashed lines) with the inner contours, the curve is not very smooth.
	This is an artifact of using DCT based numerical simulation and is not a real effect.  
	In the larger outer contours, the boundary aberration becomes less pronounced.
	Due to the inherent periodicity of the DCT, it is performed in only one quadrant. The aberrations near the axis are aberrations that occur at the boundary. The full image is reconstructed by mirroring and abetting the mirrored images in all the four quadrants.
        }
\label{fig:milkyway_contour}
\end{figure}
The pseudo mass contours on a plane perpendicular to the galactic plane are depicted in Fig.~\ref{fig:milkyway_contour}. The contours are observed to have an elliptical shape. This means that the corresponding 3-D distribution would be an ellipsoid. 
The M31 contours are similar (not shown).
Overall, it is seen that the pseudo mass distribution is qualitatively similar to the distribution of the dark matter halo~\cite{triaxial, triaxial_tilt}, in the sense that the 3-D distribution in either case is an ellipsoid. The tilt in the ellipsoid mentioned in~\cite{triaxial_tilt}, could be due to the wrap in the Milkyway galaxy~\cite{mw_wrap}, which is not modeled in the current work.
Thus, it is possible that the effects attributed to dark matter in literature, are actually the effects of the pseudo mass, i.e., the $\frac{kc^2}{8\pi G}\phi\nabla^2\frac{1}{\phi}$ term.

\subsection{The Radial Acceleration Relation (RAR)}
\label{sec:rar}
A correlation between the radial acceleration traced by the rotation curves and the acceleration predicted from the observed distribution of baryons, was reported in Ref.~\cite{rar0, rar1}. 
The question arises as to how dark matter and baryonic matter, which are supposed to be independent, can be correlated.
However, in the proposed MBG theory, the pseudo mass is a function of the gravitational potential, $\phi$. Therefore, it is natural for the baryonic matter $\rho$, and the pseudo mass term, $\frac{kc^2}{8\pi G}\phi\nabla^2\frac{1}{\phi}$, to correlate with each other.
We now explore the RAR within the framework of the MBG theory.

\begin{figure}
\includegraphics[width = 80mm,height = 80mm]{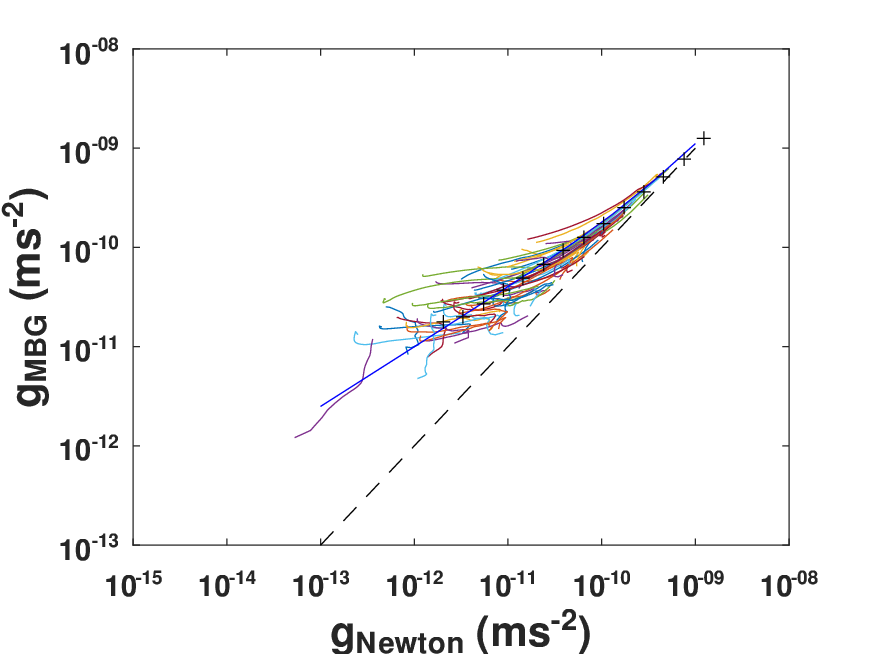}
	\caption{RAR: The colored lines are based on the MBG calculations (Eq.~\ref{eq:gal9}). The curve given by the solid line is the double power law fit to experimental data provided in Ref.~\cite{rar1}, i.e,. $y = \hat{y} \left ( 1 + \frac{x}{\hat{x}} \right )^{\alpha - \beta} \left ( \frac{x}{\hat{x}} \right )^\beta$, with $\alpha = 0.94$, $\beta = 0.6$, $\hat{x} = 2.3$, $\hat{y} = 2.6$ and $x = g_{Newton}$. The dashed line is the Newtonian limit. 
        }
\label{fig:rar}
\end{figure}

A random set of 63 galaxies of various sizes were selected from the SPARC database~\cite{rara}. 
The mass distribution of the gas in the galaxies was calculated from the velocity profile using the formula derived by Toomre~\cite{toomre}, 
and scaled based on experimentally determined total mass of gas.
Tables~\ref{tab:rar1} and~\ref{tab:rar2} present the 63 galaxies that were chosen and their calculated masses. The  $M_{\odot}/L_{\odot}$ information for the disc is also provided in tables~\ref{tab:rar1} and~\ref{tab:rar2}. The bulge $M_{\odot}/L_{\odot}$ is obtained by scaling disc $M_{\odot}/L_{\odot}$ by a factor of 1.4~\cite{rar1}.

Figure~\ref{fig:rar} depicts the plot of acceleration due to gravity obtained from the MBG theory, $g_{MBG}$, and the Newtonian gravity, $g_{Newton}$, for the selected galaxies. 
The $g_{MBG}$ and $g_{Newton}$ variables correspond to $g_{obs}$ and $g_{bar}$ respectively in Ref.~\cite{rar0,rar1}. 
 The velocity plots for the individual galaxies are provided in Figs.~\ref{fig:sparc_rot_curves_a}, ~\ref{fig:sparc_rot_curves_b} and~\ref{fig:sparc_rot_curves_c} in appendix~\ref{sec:fig_appendix}.
Based on Eq.~\ref{eq:tfr}, the same value of $k = 2.4135\times10^{-11}\sqrt{M}$ (where $M$ is the mass of the individual galaxy), provides a reasonably good fit across the mass spectrum of galaxies. 
As seen in tables~\ref{tab:rar1} and~\ref{tab:rar2} (Appendix~\ref{sec:fig_appendix}, the masses are spread across three orders of magnitude, and the proposed MBG theory, i.e. Eq.~\ref{eq:gal9} with $k= 2.4135\times10^{-11}\sqrt{M}$, is able to predict the trend.
The value of $k= 2.4135\times10^{-11}\sqrt{M}$, used for RAR, is almost same as that of $k =2.8157\times 10^{-11}\sqrt{M}$, used for the Milkyway and M31. While, one value of k is based on a fit to a wide spectrum of galaxies, the other is based on the optimal fit to two galaxies (Milkyway and M31). 
It is worth noting that the black-hole is not modeled in the SPARC dataset.
The galaxies in tables~\ref{tab:rar1} and~\ref{tab:rar2} include both gas-rich and gas-deficient galaxies.
Figure~\ref{fig:rar} indicates that the same value of $k$ is able to predict the trend across galaxies with different amounts of gas. This is consistent with the observation in Ref.~\cite{gas_galaxy}, where it was shown that the Tully Fisher relation is maintained regardless of the gas composition.

As a final note, we now discuss some of the sources of error. In galaxies such as DDO161 (Fig.~\ref{fig:DDO161}), there is a deviation at the periphery of the galaxy. One probable reason could be that, near the periphery, $\phi$ can be very small leading to inaccuracies in $\nabla^2 \frac{1}{\phi}$ calculations.
This aspect can be perceived in some other galaxies also.
While, a lot of care has been taken to maintain numerical accuracy, a superior numerical technique may improve accuracy further.

\subsection{Wide Binary Star systems}
WBS systems are a pair of stars rotating around each other.
If the separation between the stars are sufficiently large, then the gravitational acceleration becomes very weak. These are of significant interest as these systems have been used to determine as to whether dark matter theory or the MOND theory is the correct theory.
There have been mixed reports. Refs.~\cite{wbs1, wbs2} have reported that the wide binary stars deviate from the standard Newtonian gravity. However, Ref.~\cite{wbs3} has reported, that if WBS with noisy data are removed, then WBS follows Newton's laws of gravity. 

We now look at the prediction based on the MBG theory.
A binary system is not an ensemble, and is the smallest system conceivable. Thus a concept of an ensemble would not be applicable for any pure binary systems. In Eq.~\ref{eq:linenergy}, $k\langle v(t)_{rand}\rangle \rightarrow 0$. In the equivalent tensor form of Eq.~\ref{eq:linenergy}, i.e., Eq.~\ref{eq:lincomb}, the stress energy component due to thermal considerations should not play any contribution.  This can be effected by taking the degree of equilibration, $k=0$. If $k=0$, then Eq.~\ref{eq:gal9} reduces to
\begin{equation}
	\label{eq:wbs}
	\nabla^2 \phi =   4\pi G \rho,
\end{equation}
which is the Newtonian gravity.
With this, one may conclude that the proposed MBG theory predicts that any pure binary system (pure binary means that the two masses should not see any different set of interactions from the surroundings) will obey Newton's laws of gravity.

\subsection{Bullet Cluster and the UDG}
	For ultra diffuse galaxies, the grouping effect may be very small, leading to a very low value of $k$. This could lead to a very small value of the pseudo mass, as seen in the UDG, NGC 1052-DF2 and NGC 1052-DF4~\cite{DF2, DF4}. However, the ultra diffuse galaxy, Dragonfly 44~\cite{dragonfly44}, seems to have a large discrepancy between the baryonic mass and mass determined based on rotation curves. It is possible that the discrepancy is incorrect, or that another factor is in play, which may increase $k$ in UDG.

	One of the foremost pieces of evidence of dark matter is the 1E 0657-56 (Bullet cluster), where the visible matter and the matter inferred via gravitational lensing, are in different regions of space~\cite{bullet1, bullet2, bullet3, bullet4}.
	In the proposed theory, it is possible that the effect of $\square \frac{1}{\phi}$ term and the effect of visible matter, given by $\rho$, are dominant in different regions of space. 
  For instance, within a galaxy, the pseudo mass is present considerably above and below the galactic plane, where there is no stellar mass (Ref. Fig.~\ref{fig:milkyway_contour}).
  The pseudo-mass term may therefore explain the gravitational lensing phenomenon observed in 1E 0657-56.
  It is planned to analyze the bullet cluster in future. 

	\section{Conclusion}
	\label{sec:conclusion}
	We see that the proposed MBG theory, originally conceived for modeling thermal systems at the quantum mechanical level, is able to reproduce to a reasonable extent, the observed rotation curves for the Milky way and the M31 galaxy. 
	It is also able to explain the RAR relation, and consistent with WBS observations.
	Finally, in Eq.~\ref{eq:gal8}, the $\frac{1}{2s}\nabla^2 s$ term can be seen as an additional source term.
	It is possible that the presence of this term can account for the extent of the gravitational lensing phenomenon~\cite{lens1, lens2}. It is hoped to touch upon the gravitational lensing problem in future work.
	\section*{Acknowledgments}
I would like to thank Dr. Federico Lelli, for providing clarifications on the SPARC database.
I would also like to thank Dr. Fran\c{c}ois Hammer (Observatoire de Paris) for pointing me to the Milkyway rotation curves and related discussions.

\appendix
\section{The relation $k\propto \sqrt{M}$}
\label{sec:tully_appendix}
For an observer near the galactic periphery or much beyond, one may approximate the galactic mass to be centered at the origin. As such, 
one may assume a spherical symmetry as an approximation.
We now determine the solution to the asymptotic equation~\ref{eq:gal10} in spherical co-ordinates. In spherical co-ordinates,  Eq.~\ref{eq:gal10} becomes
\begin{equation}
	\label{eq:gal13}
	        \frac{1}{r^2}\frac{\partial}{\partial r} r^2 \frac{\partial \phi}{\partial r} \approx \frac{kc^2}{2}\frac{\phi}{r^2}\frac{\partial}{\partial r} r^2 \frac{\partial}{\partial r}\frac{1}{\phi}.
\end{equation}
Substitute $r = e^t$, and subsequently, assign $\zeta(\phi) = \frac{\partial \phi}{\partial t}$. This gives
\begin{equation}
\label{eq:asym2}
\zeta' - \frac{2K}{\phi^2 + K\phi}\zeta = -1,
\end{equation}
where $K = \frac{kc^2}{2}$. The integration factor for this is
\begin{equation}
\label{eq:asym3}
	f(\phi) = \left ( \frac{K + \phi}{\phi} \right )^2.
\end{equation}
Multiplying the integration factor on both sides of Eq.~\ref{eq:asym2}, and solving, we get
\begin{equation}
\label{eq:asym4}
	f(\phi) \frac{\partial \phi}{\partial t} = -\int f(\phi) d\phi.
\end{equation}
After solving, we get:
\begin{equation}
\label{eq:asym5}
	-\int f(\phi) d\phi = \frac{K^2}{\phi} - 2K\ln(|\phi|) - \phi - c_2.
\end{equation}
One may again rewrite Eq.~\ref{eq:asym4} as:
\begin{equation}
\label{eq:asym6}
	\frac{f(\phi) d\phi}{\int f(\phi) d\phi} = -dt.
\end{equation}
Solving the integration and reverse substituting $t = \ln(r)$,
\begin{equation}
\label{eq:asym7}
	\int f(\phi) d\phi = -\frac{c_1}{r}.
\end{equation}
Comparing Eqs.~\ref{eq:asym5} and~\ref{eq:asym7}, we finally obtain:
\begin{equation}
\label{eq:asym8}
	\frac{K^2}{\phi} - 2K\ln(|\phi|) - \phi  =  \frac{c_1}{r} + c_2,
\end{equation}
where, $c_1$ and $c_2$ are constants of integration. 

We now, simplify Eq.~\ref{eq:asym8}, and bring out the dominant factors that relate to $K$, in the region where Eq.~\ref{eq:gal10} is applicable, i.e.,  when $r$ is large.

For large $r$, $\frac{c_1}{r} - c_2 \approx c_2$. The gravitational potential $\phi$ would also tend to be a small value, i.e. $\phi \rightarrow \phi_0$, where $\phi_0$ is small. In Newtonian gravity, $\phi_0$ is normally taken as 0. However, in this theory, we shall see that a relation exists between $K$, $c_2$ and $\phi_0$, which prevents $\phi_0$ or $c_2$ from being zero for a non-zero $K$.

The asymptotic Eq.~\ref{eq:gal10}, is applicable only at far away distances (galactic periphery or beyond), where $\phi$ is very small. Hence, the limiting value as $r\rightarrow \infty$, i.e., $\phi_0$, should be very small. 
Since, 	$\lim_{\phi_0 \rightarrow 0} \phi_0 \ln(\phi_0) = 0$,
for a small $\phi_0$, one may approximate:
\begin{equation}
\label{eq:asym9}
K^2 - 2K\phi\ln(|\phi_0|) \approx K^2 . 
\end{equation}
Then, Eq.~\ref{eq:asym8} reduces to
\begin{equation}
\label{eq:asym10}
	K^2  \approx  c_2\phi_0  + \phi_0^2.
\end{equation}
For a given non-zero $K$, if $\phi_0$ is very small, then 
$c_2$ has to be large enough to satisfy Eq.~\ref{eq:asym10}
and $c_2\phi_0$ becomes dominant. 
In other words, $c_2 \gg \phi_0$.
Subsequently,
\begin{equation}
\label{eq:asym11}
	K  \approx  \pm \sqrt{c_2\phi_0}.
\end{equation}
Since $K$ is positive, 
\begin{equation}
\label{eq:asym12}
	K  =  \sqrt{c_2\phi_0}.
\end{equation}
As a check, we substitute  $K  =  \sqrt{c_2\phi_0}$ in E.~\ref{eq:asym9} to obtain,
\begin{multline}
\label{eq:asym13}
	K^2 - k\phi_0 \ln(|\phi_0|) = 
	c_2\phi_0 - 2\sqrt{c_2}\phi_0^{3/2}\ln(|\phi|_0) \\
	= 2\sqrt{c_2}\phi_0 \left [ \frac{\sqrt{c_2}}{2} - \sqrt{\phi_0} \ln (|\phi_0|) \right ] \\
	\approx c\phi_0 = K^2,
\end{multline}
thus validating Eq.~\ref{eq:asym10}.
The last line in Eq.~\ref{eq:asym13}, makes use of the fact that, since $\lim_{\phi_0 \rightarrow 0} \sqrt{\phi_0} \ln(\phi_0) = 0$,  $\sqrt{\phi_0} \ln (|\phi_0|)$, would become negligible for very small $\phi_0$. 

To determine the relation of $\phi_0$ with $M$, we start with Newtoniant gravity as the starting point and iterate.
To begin with, if we substitute $\phi_0 \propto M$ in Eq.~\ref{eq:asym12}, we get $K  \propto  \sqrt{M}$. 
As one iterates, the $\phi\nabla^2\frac{1}{\phi}$ term would apply corrections to $\phi$. If, $K$ is of the order of $\sqrt{M}$, the first order corrections to $\phi$ would be of the order $\sqrt{M}$, i.e., $\phi_0 \sim \bigO(M) + \bigO(\sqrt(M)) + ...$. For large $M$, if $M^1$ is dominant, one can ignore these corrections.
Then, as an approximation, substituting $\phi_0 \propto M$, in Eq.~\ref{eq:asym12}, we get 
\begin{equation}
\label{eq:asym14}
	K  \propto  \sqrt{M}.
\end{equation}
Substituting $K\propto \sqrt{M}$ (and hence, $k \propto \sqrt{M}$), in Eq.\ref{eq:gal12}, we reproduce the Tully Fisher relation. 

\section{Rotation curve Figures and tables.}
\label{sec:fig_appendix}
\begin{figure*}
     \begin{subfigure}[b]{0.5\columnwidth}
         \centering
         \includegraphics[width=\columnwidth]{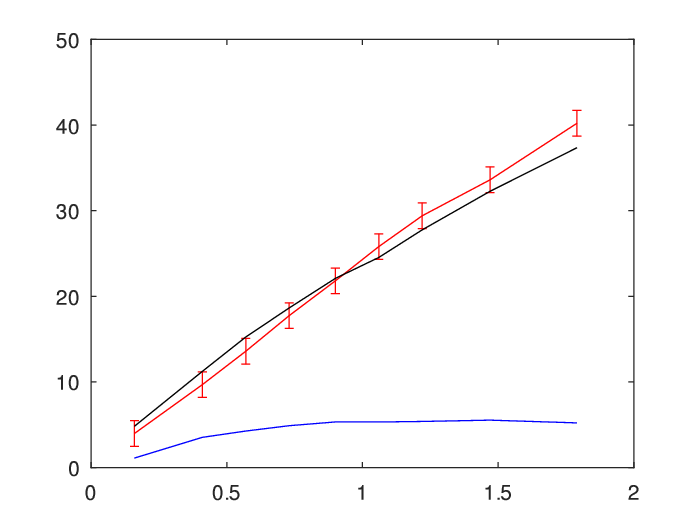}
         \caption{ CamB}
         \label{fig:CamB}
     \end{subfigure}
     \hfill
     \begin{subfigure}[b]{0.5\columnwidth}
         \centering
         \includegraphics[width=\columnwidth]{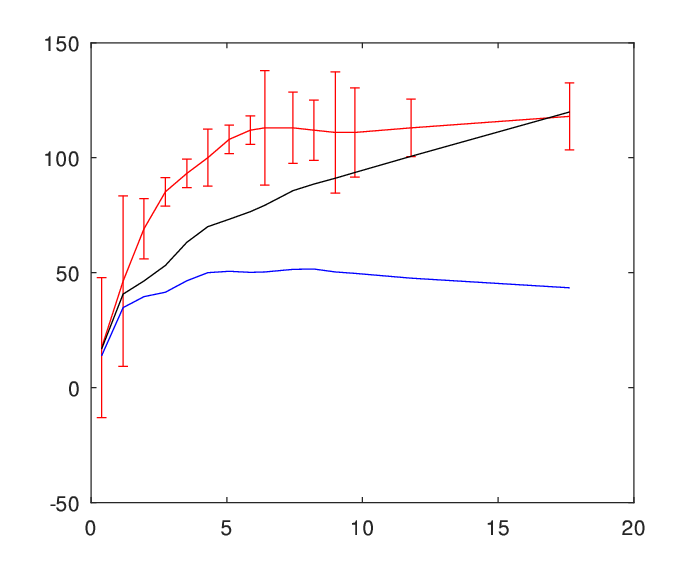}
         \caption{ F568-V1}
         \label{fig:F568-V1}
     \end{subfigure}
     \hfill
     \begin{subfigure}[b]{0.5\columnwidth}
         \centering
         \includegraphics[width=\columnwidth]{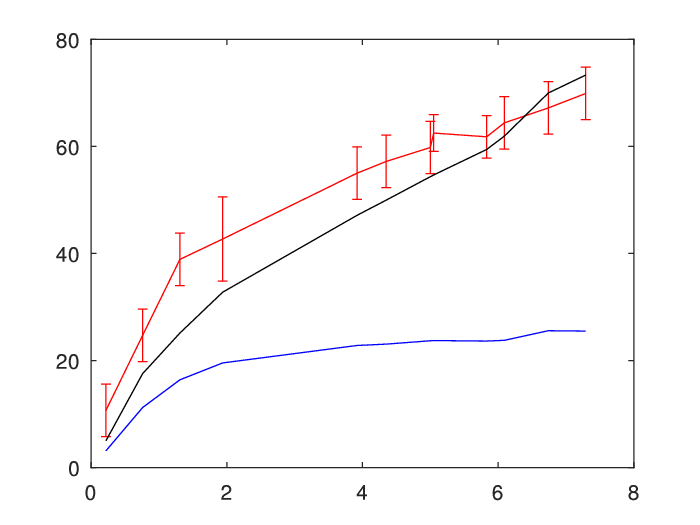}
         \caption{ F583-4}
         \label{fig:F583-4}
     \end{subfigure}
     \hfill
     \begin{subfigure}[b]{0.5\columnwidth}
         \centering
         \includegraphics[width=\columnwidth]{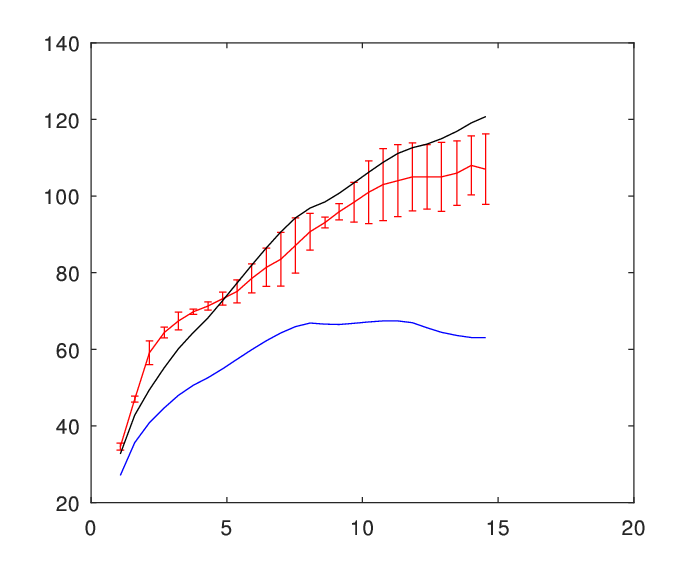}
         \caption{ NGC0247}
         \label{fig:NGC0247}
     \end{subfigure}
     \hfill
     \begin{subfigure}[b]{0.5\columnwidth}
         \centering
         \includegraphics[width=\columnwidth]{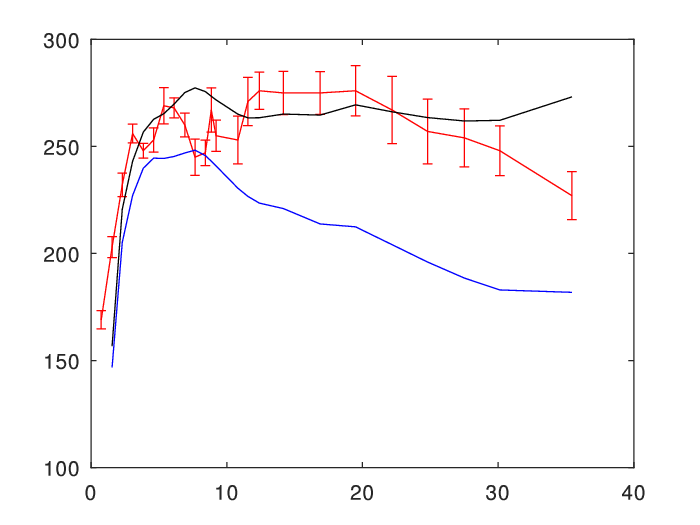}
         \caption{ NGC2955}
         \label{fig:NGC2955}
     \end{subfigure}
     \hfill
     \begin{subfigure}[b]{0.5\columnwidth}
         \centering
         \includegraphics[width=\columnwidth]{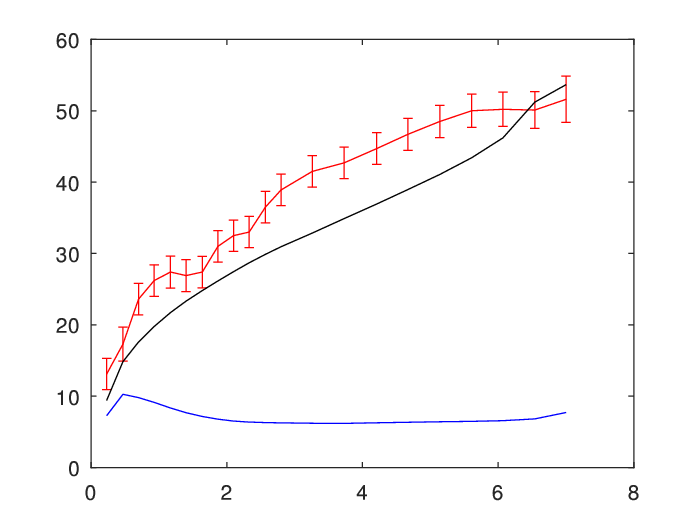}
         \caption{ NGC3741}
         \label{fig:NGC3741}
     \end{subfigure}
     \hfill
     \begin{subfigure}[b]{0.5\columnwidth}
         \centering
         \includegraphics[width=\columnwidth]{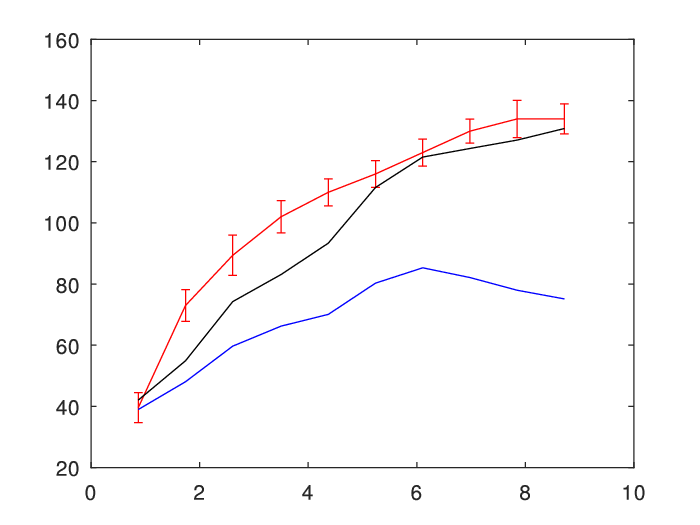}
         \caption{ NGC3972}
         \label{fig:NGC3972}
     \end{subfigure}
     \hfill
     \begin{subfigure}[b]{0.5\columnwidth}
         \centering
         \includegraphics[width=\columnwidth]{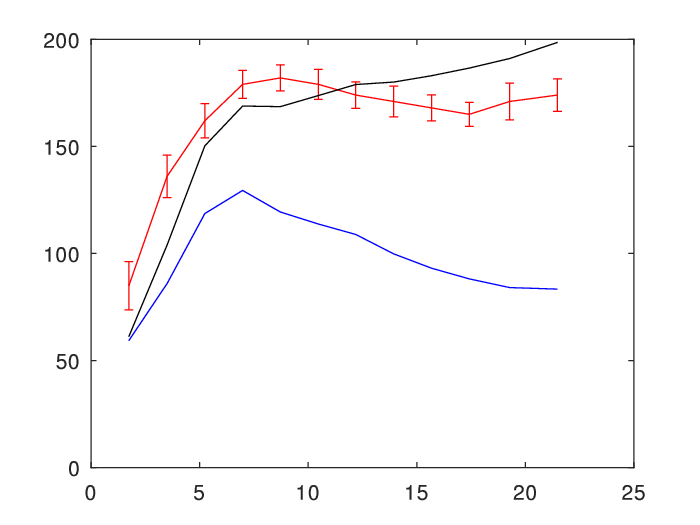}
         \caption{ NGC4088}
         \label{fig:NGC4088}
     \end{subfigure}
     \hfill
     \begin{subfigure}[b]{0.5\columnwidth}
         \centering
         \includegraphics[width=\columnwidth]{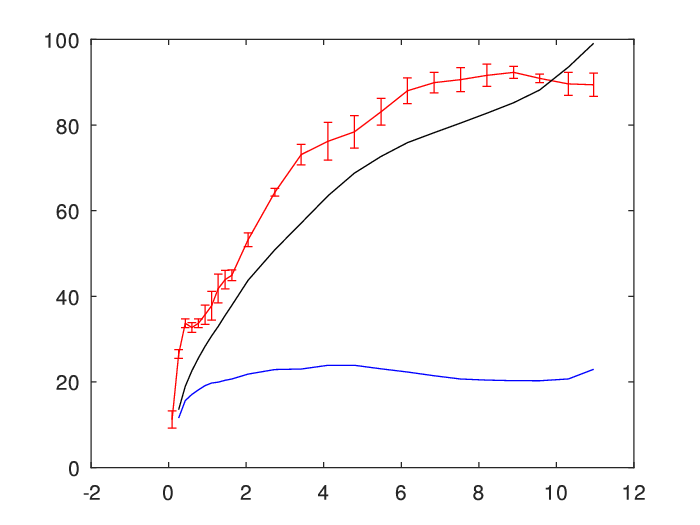}
         \caption{ NGC5585}
         \label{fig:NGC5585}
     \end{subfigure}
     \hfill
     \begin{subfigure}[b]{0.5\columnwidth}
         \centering
         \includegraphics[width=\columnwidth]{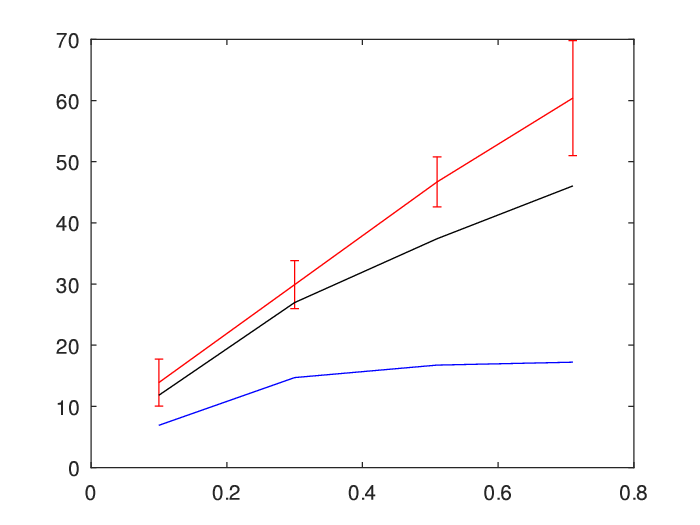}
         \caption{ NGC6789}
         \label{fig:NGC6789}
     \end{subfigure}
     \hfill
     \begin{subfigure}[b]{0.5\columnwidth}
         \centering
         \includegraphics[width=\columnwidth]{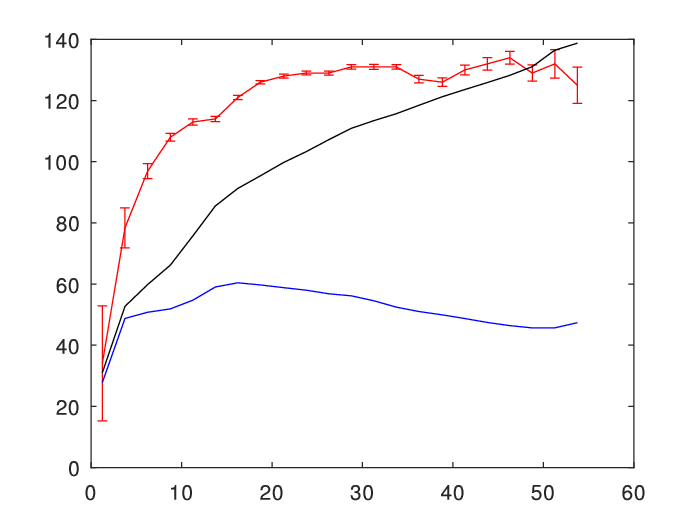}
         \caption{ UGC00128}
         \label{fig:UGC00128}
     \end{subfigure}
     \hfill
     \begin{subfigure}[b]{0.5\columnwidth}
         \centering
         \includegraphics[width=\columnwidth]{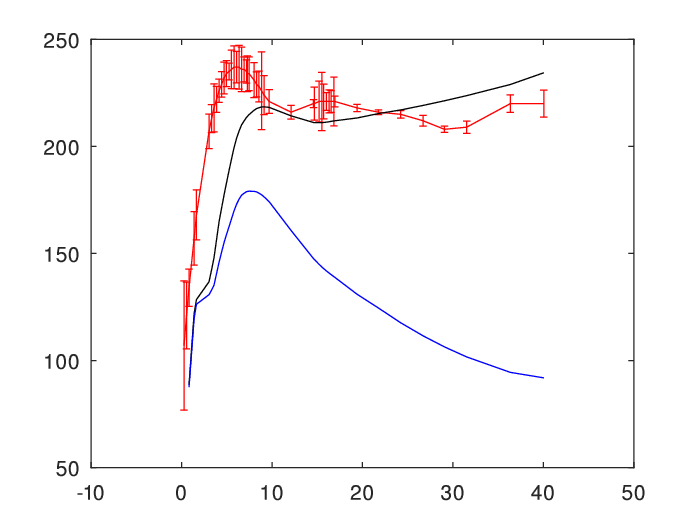}
         \caption{ UGC03205}
         \label{fig:UGC03205}
     \end{subfigure}
     \hfill
     \begin{subfigure}[b]{0.5\columnwidth}
         \centering
         \includegraphics[width=\columnwidth]{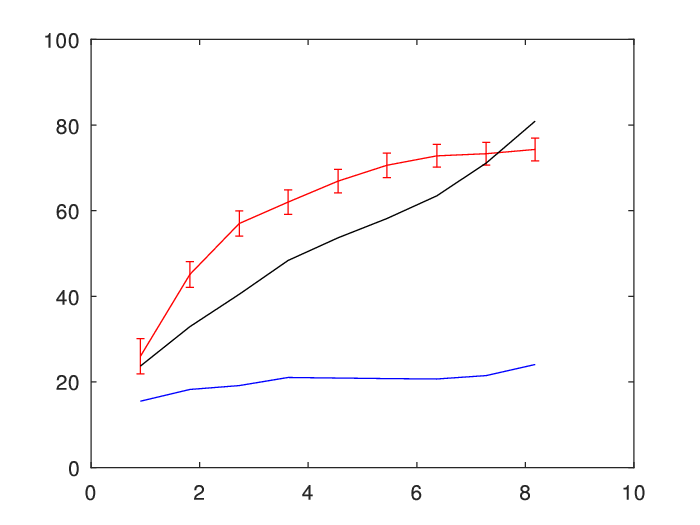}
         \caption{ UGC04499}
         \label{fig:UGC04499}
     \end{subfigure}
     \hfill
     \begin{subfigure}[b]{0.5\columnwidth}
         \centering
         \includegraphics[width=\columnwidth]{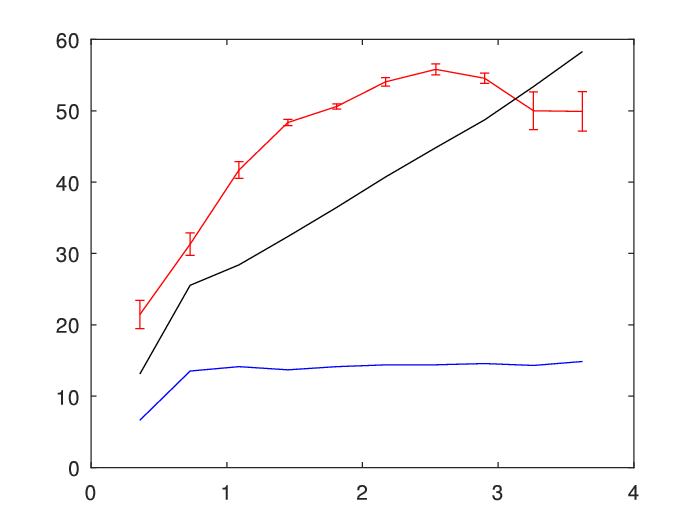}
         \caption{ UGC05764}
         \label{fig:UGC05764}
     \end{subfigure}
     \hfill
     \begin{subfigure}[b]{0.5\columnwidth}
         \centering
         \includegraphics[width=\columnwidth]{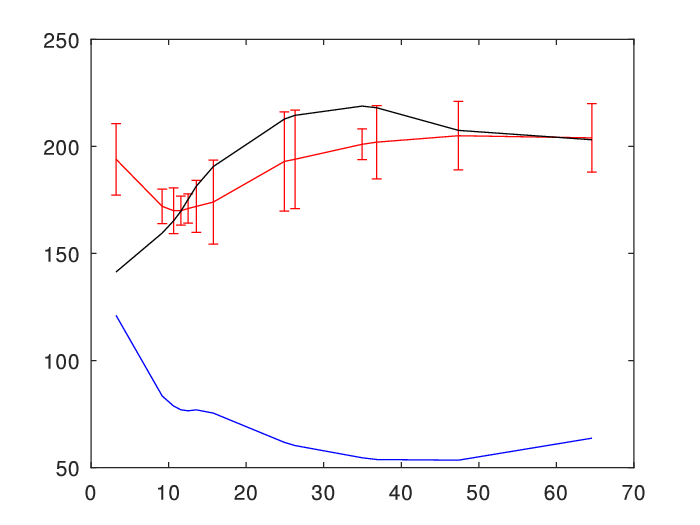}
         \caption{ UGC06614}
         \label{fig:UGC06614}
     \end{subfigure}
     \hfill
     \begin{subfigure}[b]{0.5\columnwidth}
         \centering
         \includegraphics[width=\columnwidth]{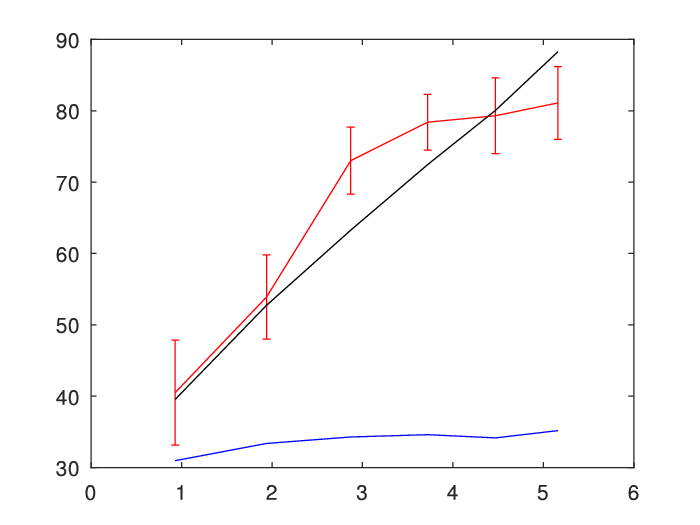}
         \caption{ UGC06923}
         \label{fig:UGC06923}
     \end{subfigure}
     \hfill
     \begin{subfigure}[b]{0.5\columnwidth}
         \centering
         \includegraphics[width=\columnwidth]{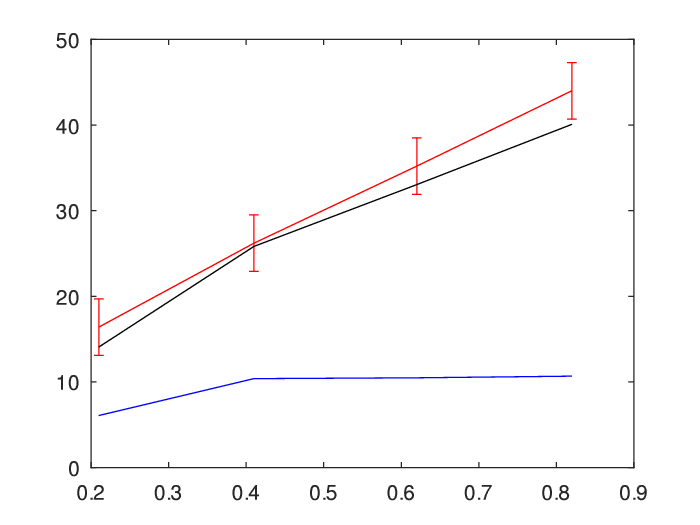}
         \caption{ UGC07232}
         \label{fig:UGC07232}
     \end{subfigure}
     \hfill
     \begin{subfigure}[b]{0.5\columnwidth}
         \centering
         \includegraphics[width=\columnwidth]{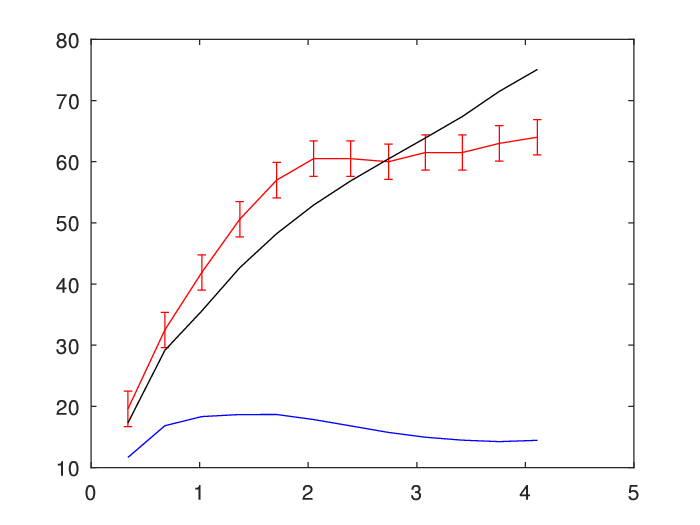}
         \caption{ UGC07603}
         \label{fig:UGC07603}
     \end{subfigure}
     \hfill
     \begin{subfigure}[b]{0.5\columnwidth}
         \centering
         \includegraphics[width=\columnwidth]{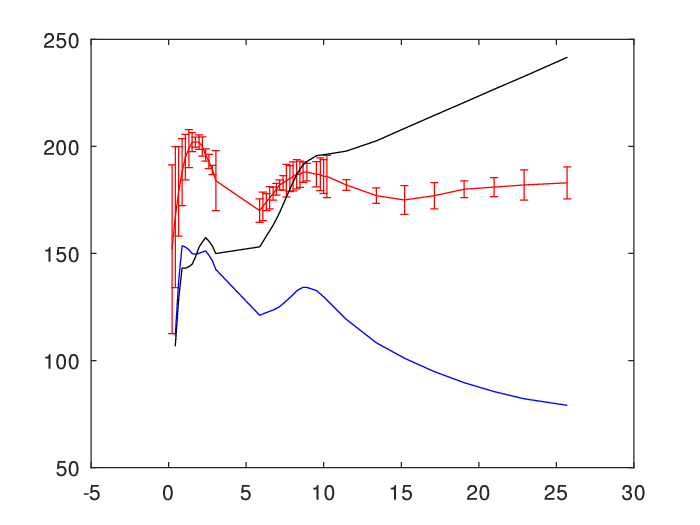}
         \caption{ UGC08699}
         \label{fig:UGC08699}
     \end{subfigure}
     \hfill
     \begin{subfigure}[b]{0.5\columnwidth}
         \centering
         \includegraphics[width=\columnwidth]{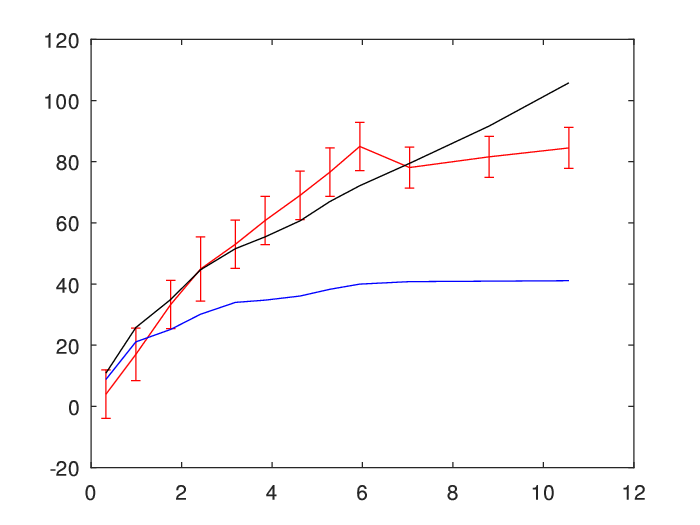}
         \caption{ UGC11557}
         \label{fig:UGC11557}
     \end{subfigure}
     \hfill

        \caption{Rotation Curves. The red curve is the experimentally observed curve, the black curve is calculated using MBG, and the blue curve is Newtonian.}
        \label{fig:sparc_rot_curves_a}
\end{figure*}

\begin{figure*}

     \begin{subfigure}[b]{0.5\columnwidth}
         \centering
         \includegraphics[width=\columnwidth]{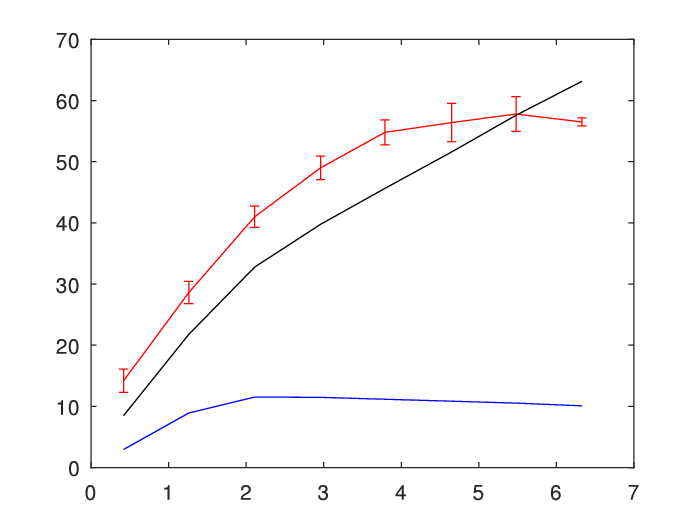}
         \caption{ UGCA442}
         \label{fig:UGCA442}
     \end{subfigure}
     \hfill
     \begin{subfigure}[b]{0.5\columnwidth}
         \centering
         \includegraphics[width=\columnwidth]{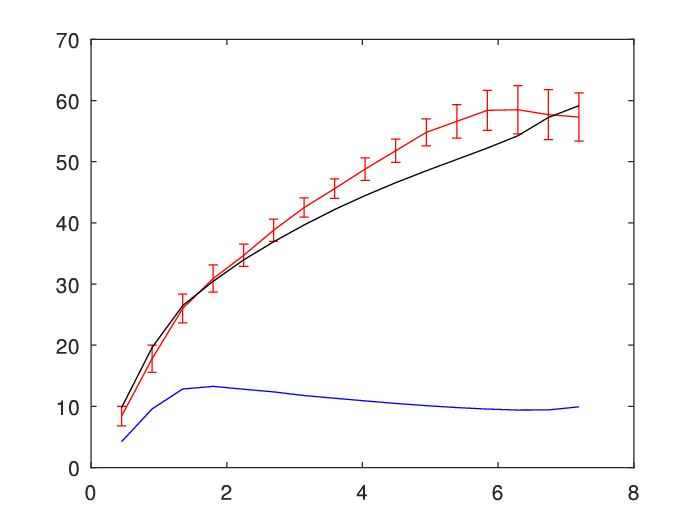}
         \caption{ D631-7}
         \label{fig:D631-7}
     \end{subfigure}
     \hfill
     \begin{subfigure}[b]{0.5\columnwidth}
         \centering
         \includegraphics[width=\columnwidth]{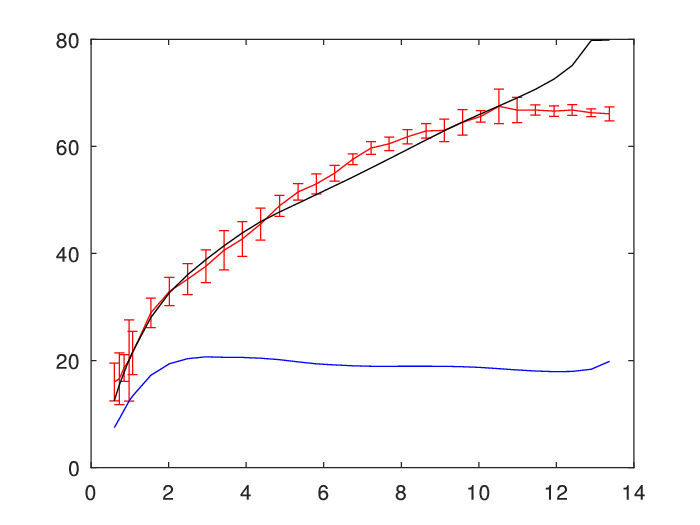}
         \caption{ DDO161}
         \label{fig:DDO161}
     \end{subfigure}
     \hfill
     \begin{subfigure}[b]{0.5\columnwidth}
         \centering
         \includegraphics[width=\columnwidth]{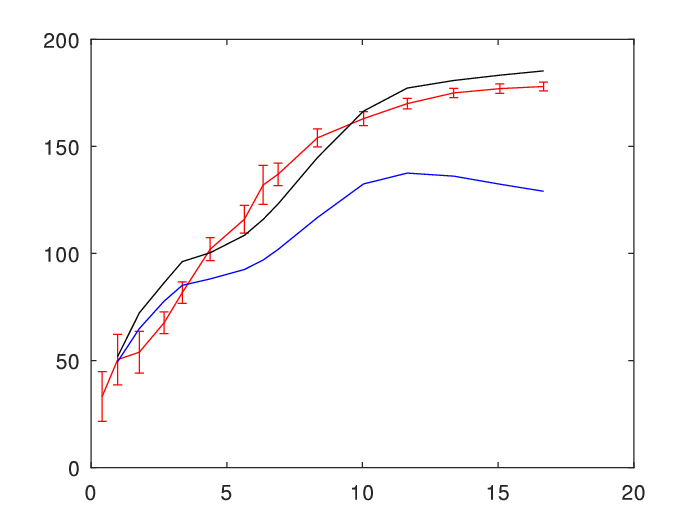}
         \caption{ ESO079-G014}
         \label{fig:ESO079-G014}
     \end{subfigure}
     \hfill
     \begin{subfigure}[b]{0.5\columnwidth}
         \centering
         \includegraphics[width=\columnwidth]{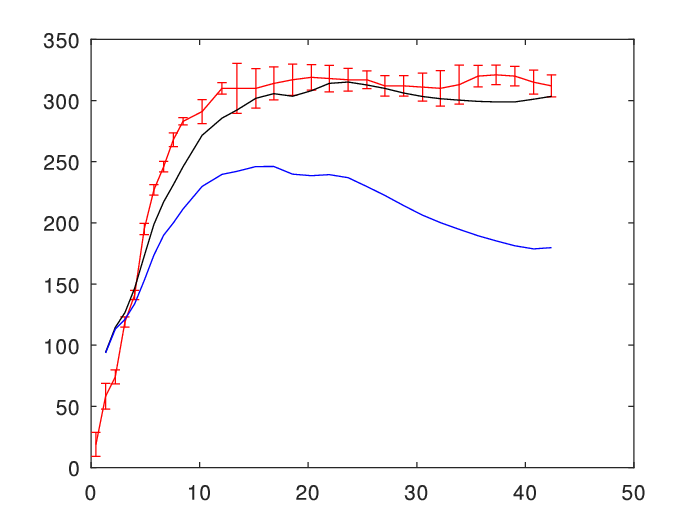}
         \caption{ ESO563-G021}
         \label{fig:ESO563-G021}
     \end{subfigure}
     \hfill
     \begin{subfigure}[b]{0.5\columnwidth}
         \centering
         \includegraphics[width=\columnwidth]{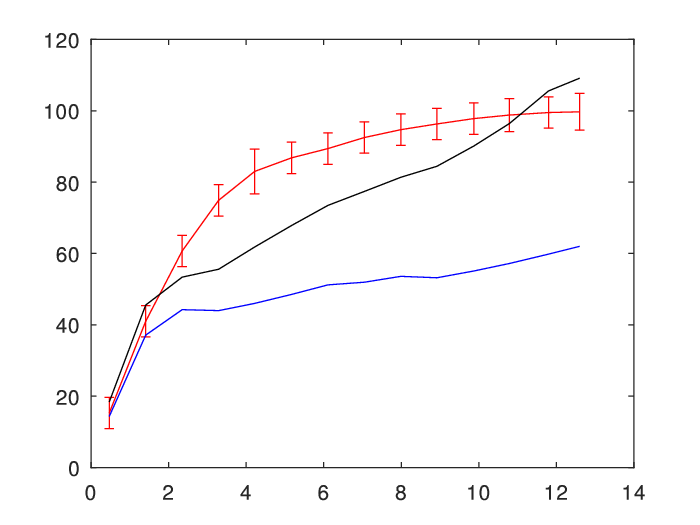}
         \caption{ F574-1}
         \label{fig:F574-1}
     \end{subfigure}
     \hfill
     \begin{subfigure}[b]{0.5\columnwidth}
         \centering
         \includegraphics[width=\columnwidth]{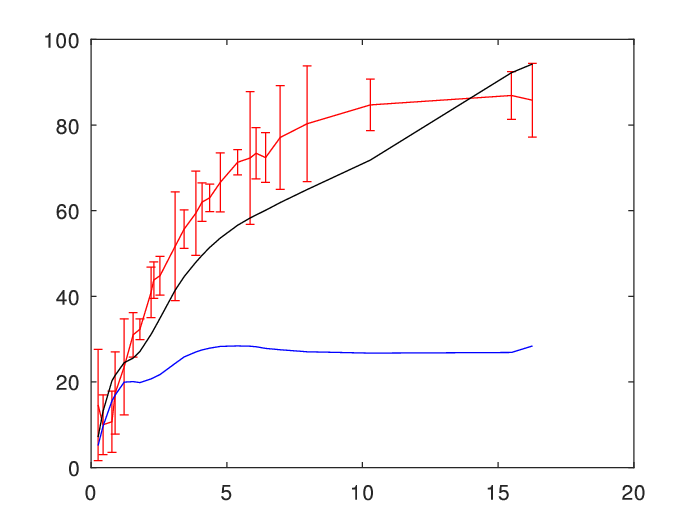}
         \caption{ F583-1}
         \label{fig:F583-1}
     \end{subfigure}
     \hfill
     \begin{subfigure}[b]{0.5\columnwidth}
         \centering
         \includegraphics[width=\columnwidth]{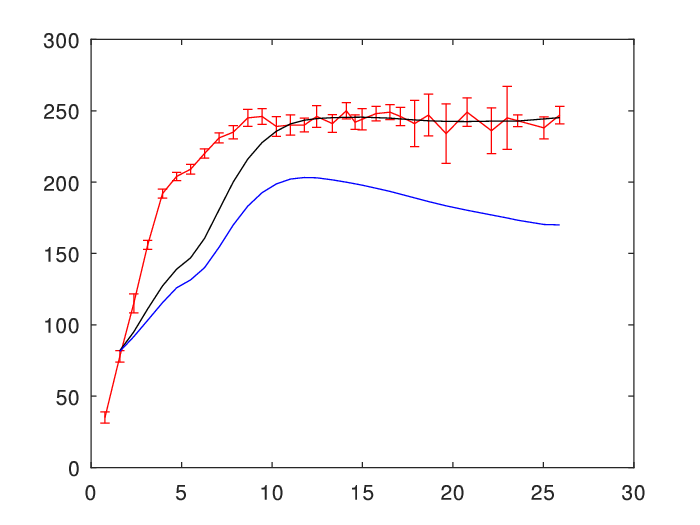}
         \caption{ IC4202}
         \label{fig:IC4202}
     \end{subfigure}
     \hfill
     \begin{subfigure}[b]{0.5\columnwidth}
         \centering
         \includegraphics[width=\columnwidth]{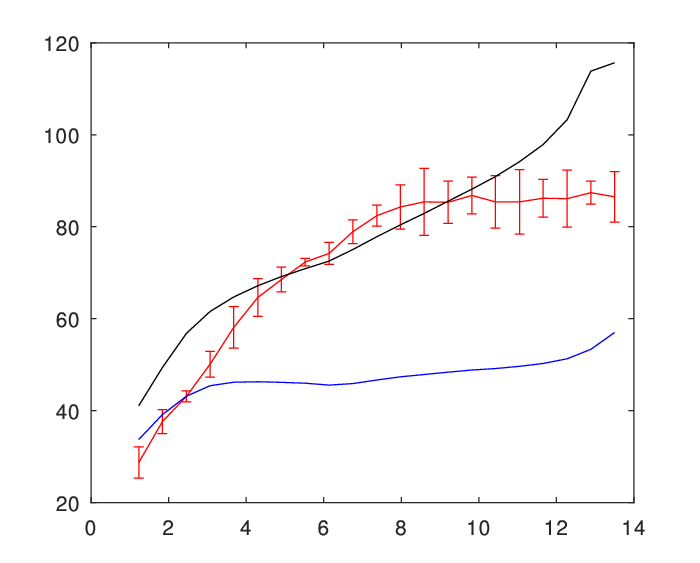}
         \caption{ NGC0055}
         \label{fig:NGC0055}
     \end{subfigure}
     \hfill
     \begin{subfigure}[b]{0.5\columnwidth}
         \centering
         \includegraphics[width=\columnwidth]{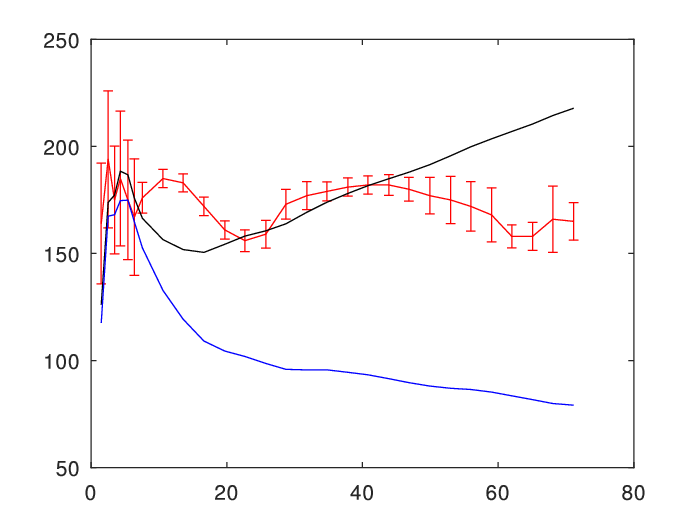}
         \caption{ NGC0289}
         \label{fig:NGC0289}
     \end{subfigure}
     \hfill
     \begin{subfigure}[b]{0.5\columnwidth}
         \centering
         \includegraphics[width=\columnwidth]{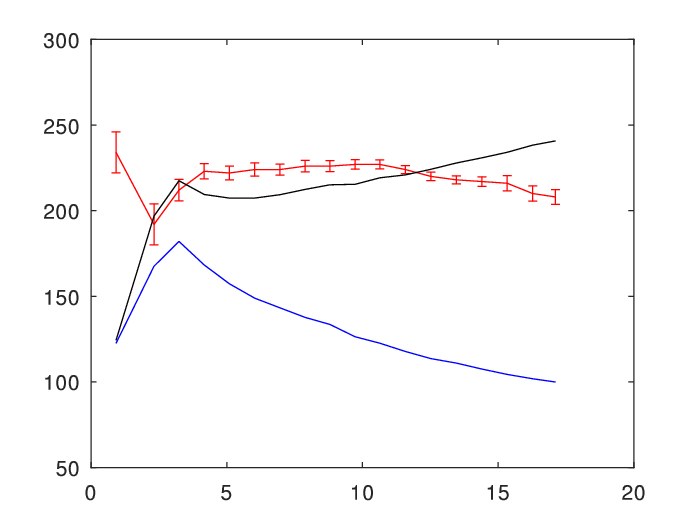}
         \caption{ NGC0891}
         \label{fig:NGC0891}
     \end{subfigure}
     \hfill
     \begin{subfigure}[b]{0.5\columnwidth}
         \centering
         \includegraphics[width=\columnwidth]{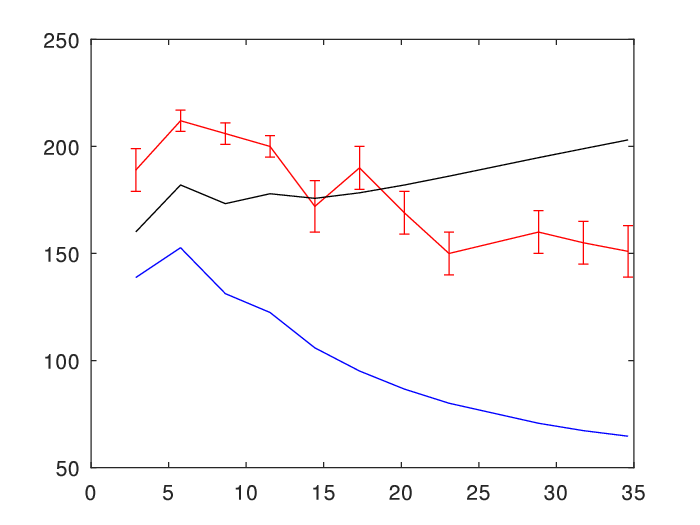}
         \caption{ NGC2683}
         \label{fig:NGC2683}
     \end{subfigure}
     \hfill
     \begin{subfigure}[b]{0.5\columnwidth}
         \centering
         \includegraphics[width=\columnwidth]{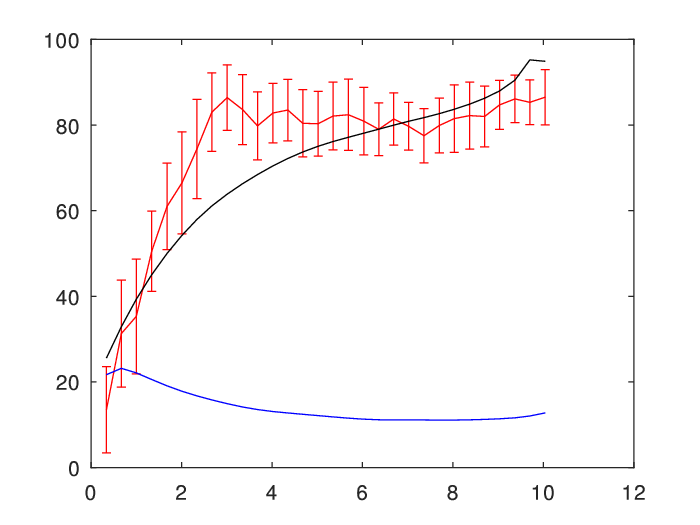}
         \caption{ NGC2915}
         \label{fig:NGC2915}
     \end{subfigure}
     \hfill
     \begin{subfigure}[b]{0.5\columnwidth}
         \centering
         \includegraphics[width=\columnwidth]{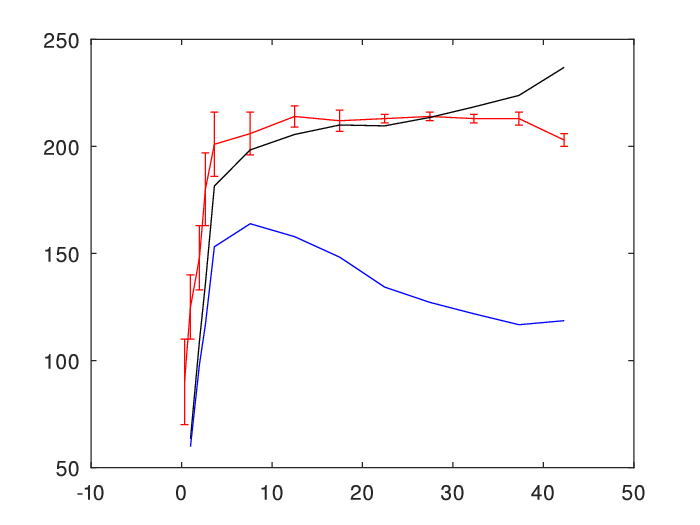}
         \caption{ NGC2998}
         \label{fig:NGC2998}
     \end{subfigure}
     \hfill
     \begin{subfigure}[b]{0.5\columnwidth}
         \centering
         \includegraphics[width=\columnwidth]{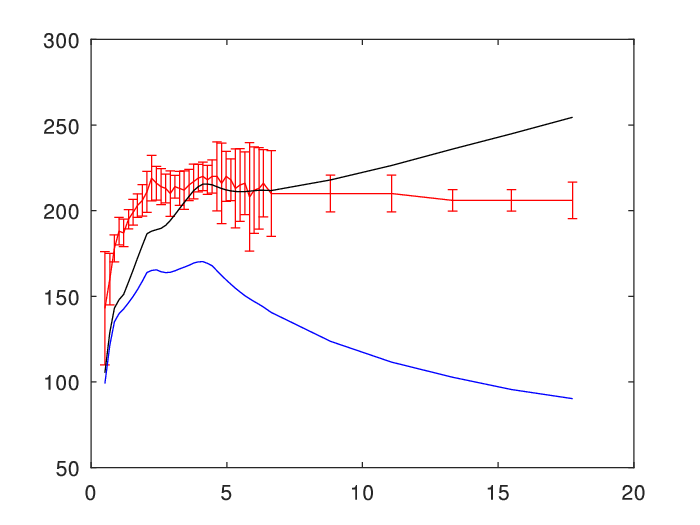}
         \caption{ NGC3521}
         \label{fig:NGC3521}
     \end{subfigure}
     \hfill
     \begin{subfigure}[b]{0.5\columnwidth}
         \centering
         \includegraphics[width=\columnwidth]{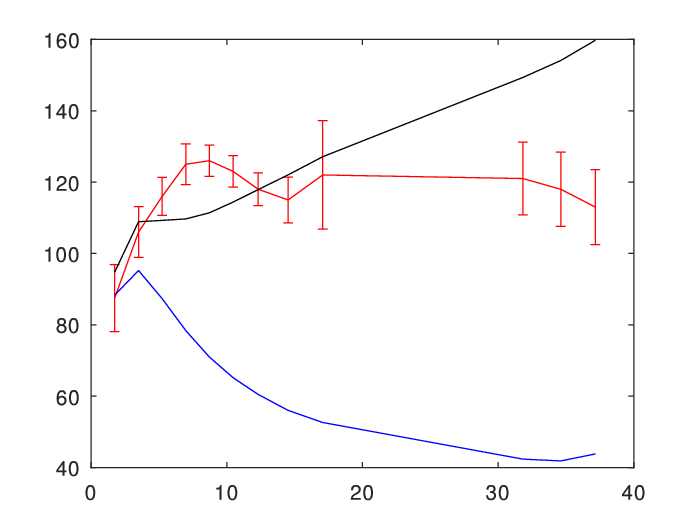}
         \caption{ NGC3769}
         \label{fig:NGC3769}
     \end{subfigure}
     \hfill
     \begin{subfigure}[b]{0.5\columnwidth}
         \centering
         \includegraphics[width=\columnwidth]{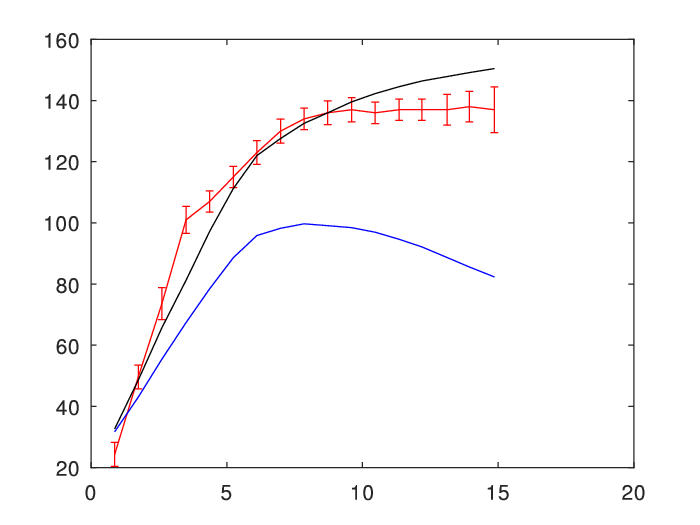}
         \caption{ NGC3917}
         \label{fig:NGC3917}
     \end{subfigure}
     \hfill
     \begin{subfigure}[b]{0.5\columnwidth}
         \centering
         \includegraphics[width=\columnwidth]{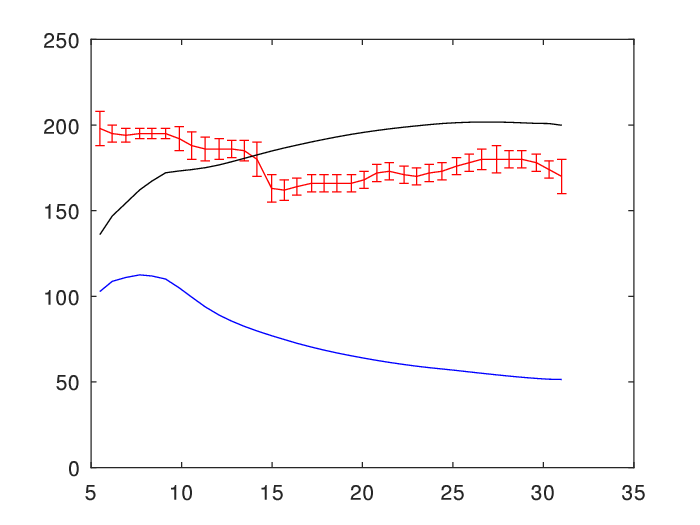}
         \caption{ NGC4013}
         \label{fig:NGC4013}
     \end{subfigure}
     \hfill
     \begin{subfigure}[b]{0.5\columnwidth}
         \centering
         \includegraphics[width=\columnwidth]{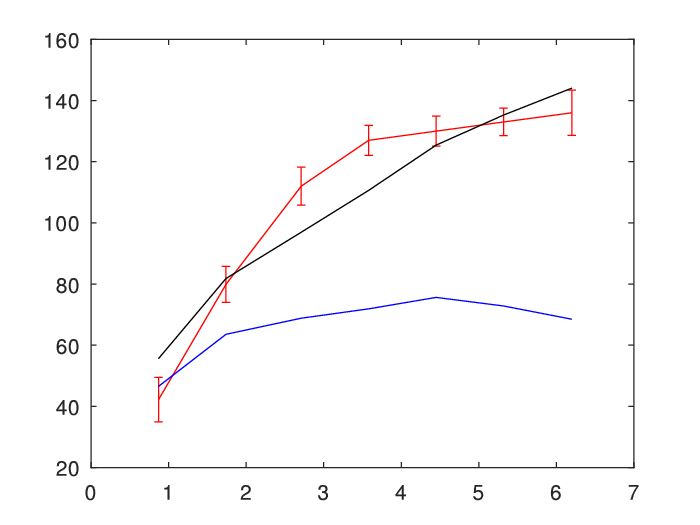}
         \caption{ NGC4085}
         \label{fig:NGC4085}
     \end{subfigure}
     \hfill
     \begin{subfigure}[b]{0.5\columnwidth}
         \centering
         \includegraphics[width=\columnwidth]{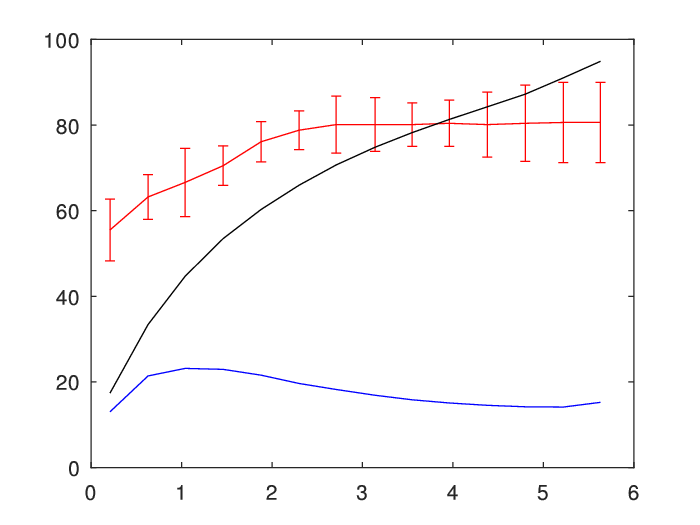}
         \caption{ NGC4214}
         \label{fig:NGC4214}
     \end{subfigure}
     \hfill

        \caption{Rotation Curves. The red curve is the experimentally observed curve, the black curve is calculated using MBG, and the blue curve is Newtonian.}
        \label{fig:sparc_rot_curves_b}
\end{figure*}

\begin{figure*}

     \begin{subfigure}[b]{0.5\columnwidth}
         \centering
         \includegraphics[width=\columnwidth]{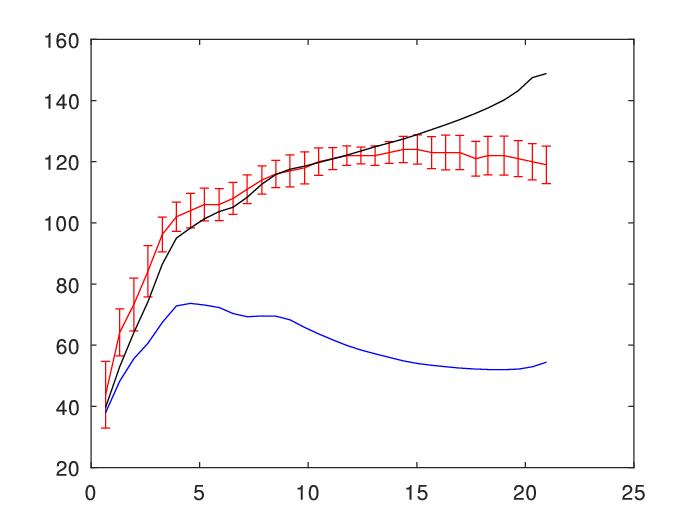}
         \caption{ NGC4559}
         \label{fig:NGC4559}
     \end{subfigure}
     \hfill
     \begin{subfigure}[b]{0.5\columnwidth}
         \centering
         \includegraphics[width=\columnwidth]{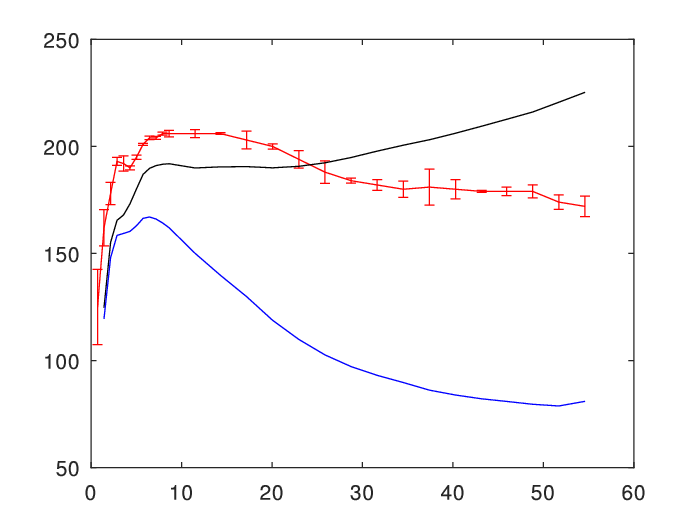}
         \caption{ NGC5055}
         \label{fig:NGC5055}
     \end{subfigure}
     \hfill
     \begin{subfigure}[b]{0.5\columnwidth}
         \centering
         \includegraphics[width=\columnwidth]{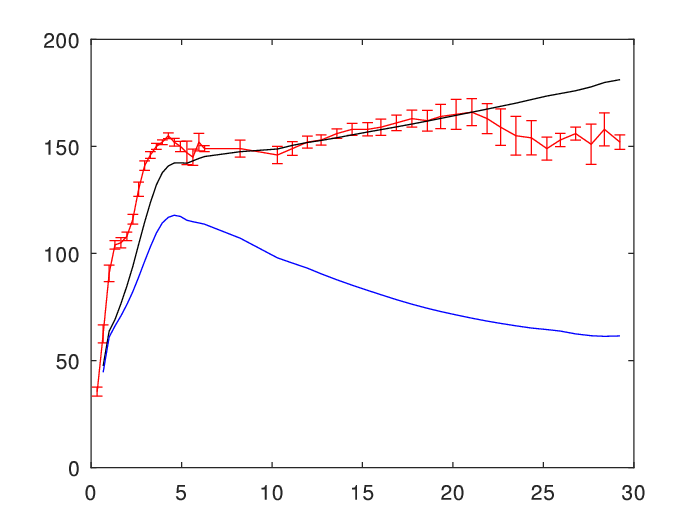}
         \caption{ NGC6015}
         \label{fig:NGC6015}
     \end{subfigure}
     \hfill
     \begin{subfigure}[b]{0.5\columnwidth}
         \centering
         \includegraphics[width=\columnwidth]{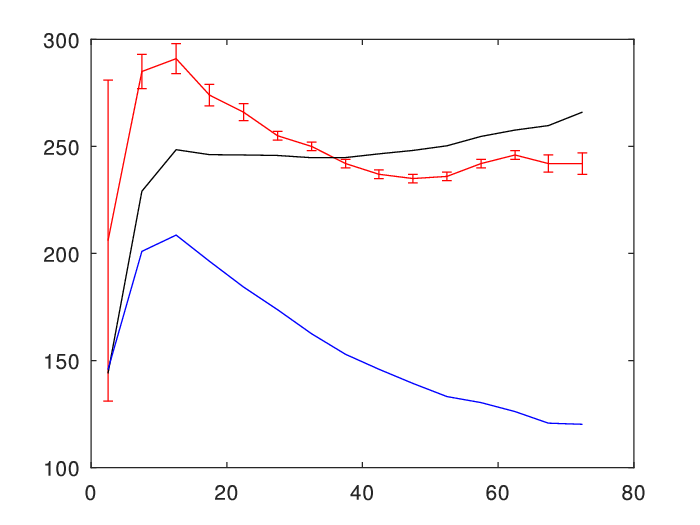}
         \caption{ NGC6674}
         \label{fig:NGC6674}
     \end{subfigure}
     \hfill
     \begin{subfigure}[b]{0.5\columnwidth}
         \centering
         \includegraphics[width=\columnwidth]{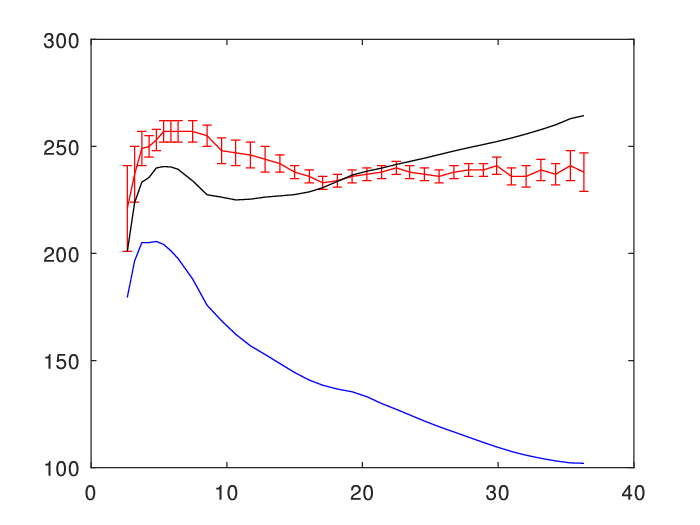}
         \caption{ NGC7331}
         \label{fig:NGC7331}
     \end{subfigure}
     \hfill
     \begin{subfigure}[b]{0.5\columnwidth}
         \centering
         \includegraphics[width=\columnwidth]{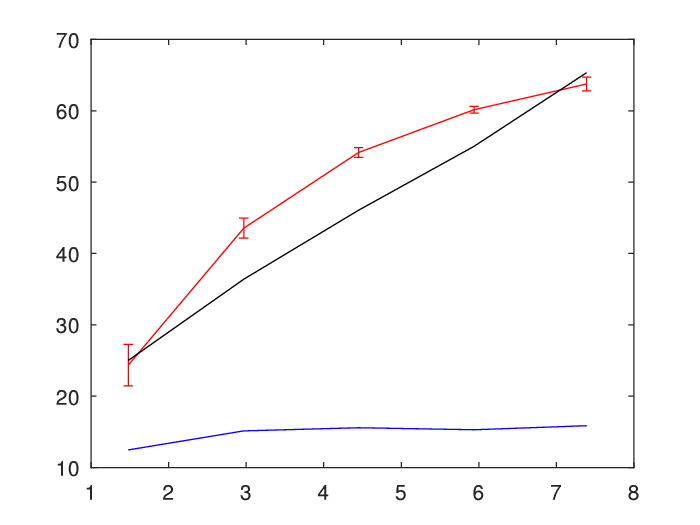}
         \caption{ UGC00891}
         \label{fig:UGC00891}
     \end{subfigure}
     \hfill
     \begin{subfigure}[b]{0.5\columnwidth}
         \centering
         \includegraphics[width=\columnwidth]{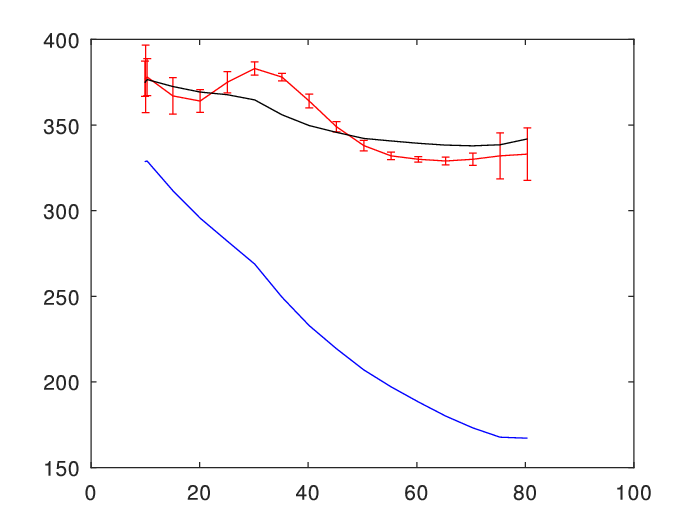}
         \caption{ UGC02487}
         \label{fig:UGC02487}
     \end{subfigure}
     \hfill
     \begin{subfigure}[b]{0.5\columnwidth}
         \centering
         \includegraphics[width=\columnwidth]{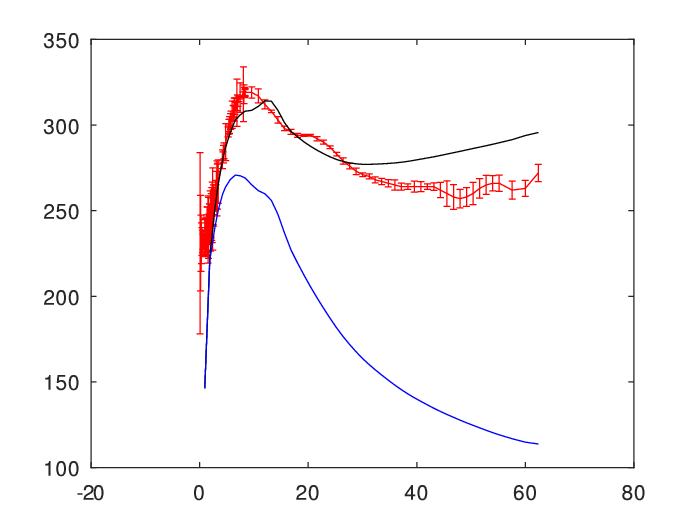}
         \caption{ UGC02953}
         \label{fig:UGC02953}
     \end{subfigure}
     \hfill

     \begin{subfigure}[b]{0.5\columnwidth}
         \centering
         \includegraphics[width=\columnwidth]{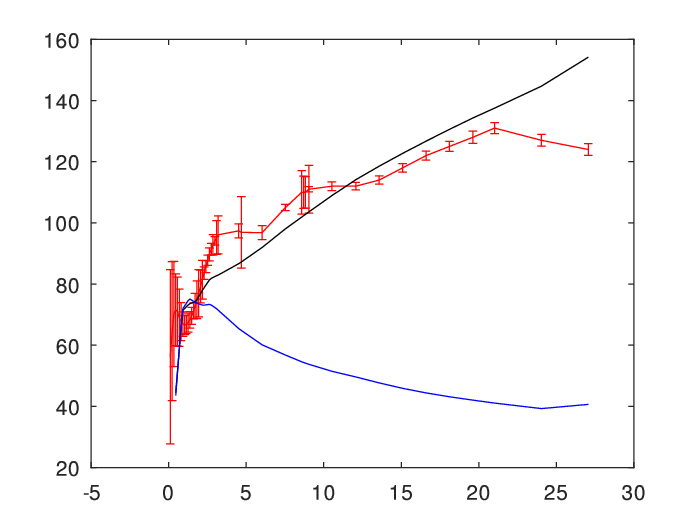}
         \caption{ UGC03580}
         \label{fig:UGC03580}
     \end{subfigure}
     \hfill
     \begin{subfigure}[b]{0.5\columnwidth}
         \centering
         \includegraphics[width=\columnwidth]{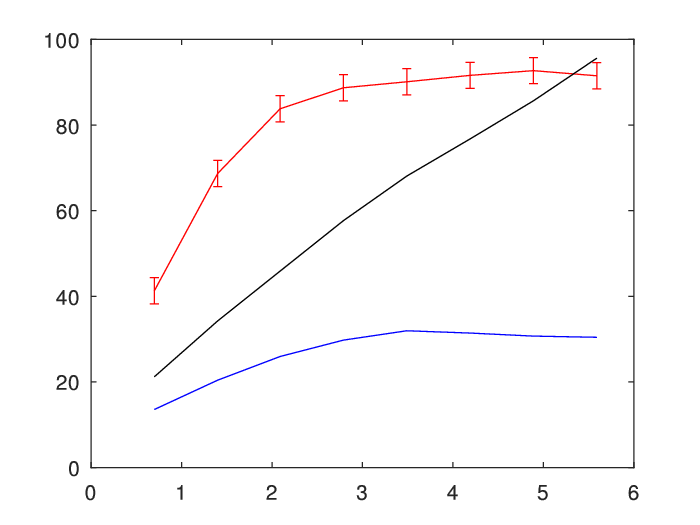}
         \caption{ UGC04325}
         \label{fig:UGC04325}
     \end{subfigure}
     \hfill
     \begin{subfigure}[b]{0.5\columnwidth}
         \centering
         \includegraphics[width=\columnwidth]{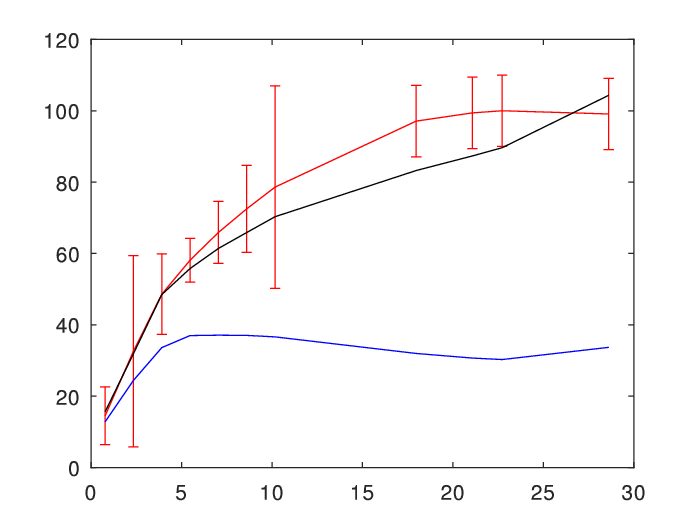}
         \caption{ UGC05005}
         \label{fig:UGC05005}
     \end{subfigure}
     \hfill
     \begin{subfigure}[b]{0.5\columnwidth}
         \centering
         \includegraphics[width=\columnwidth]{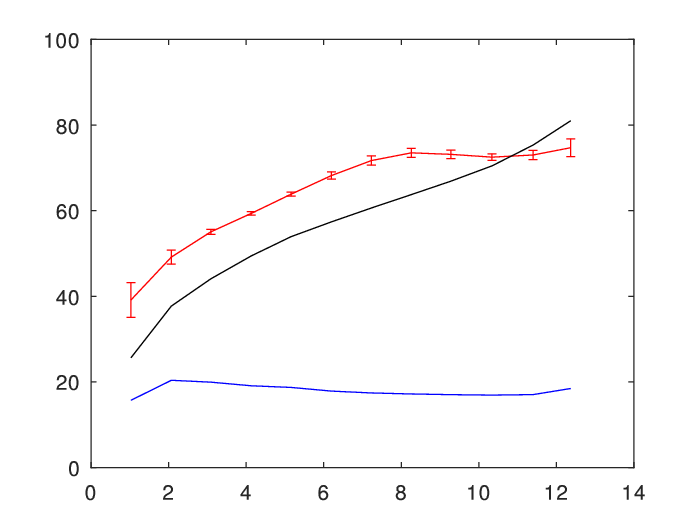}
         \caption{ UGC05716}
         \label{fig:UGC05716}
     \end{subfigure}
     \hfill
     \begin{subfigure}[b]{0.5\columnwidth}
         \centering
         \includegraphics[width=\columnwidth]{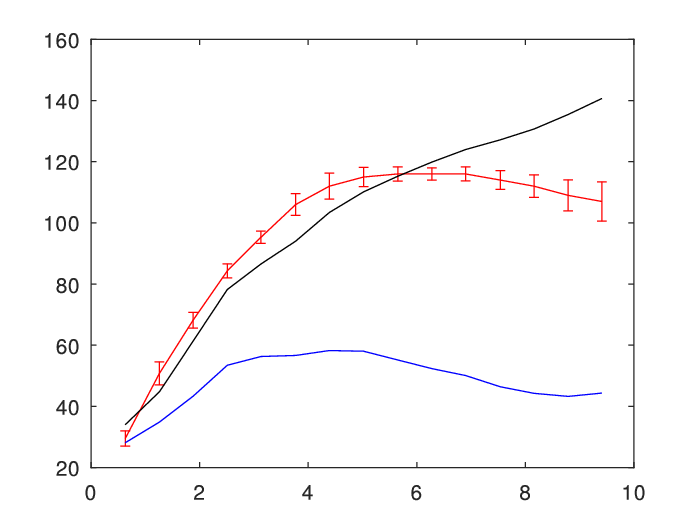}
         \caption{ UGC05986}
         \label{fig:UGC05986}
     \end{subfigure}
     \hfill
     \begin{subfigure}[b]{0.5\columnwidth}
         \centering
         \includegraphics[width=\columnwidth]{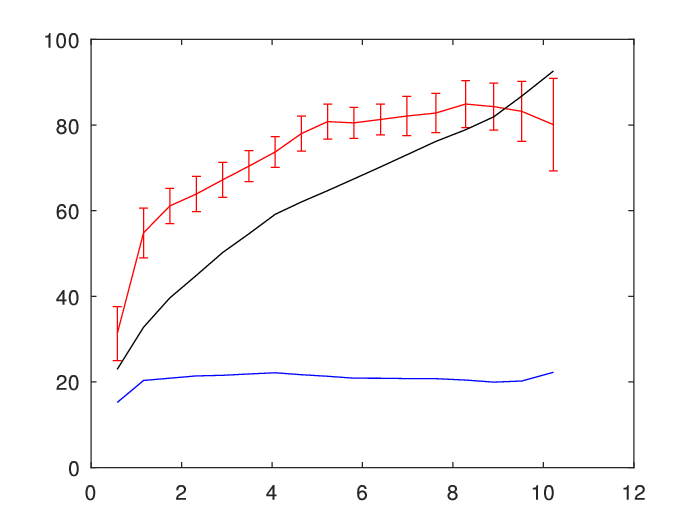}
         \caption{ UGC06446}
         \label{fig:UGC06446}
     \end{subfigure}
     \hfill
     \begin{subfigure}[b]{0.5\columnwidth}
         \centering
         \includegraphics[width=\columnwidth]{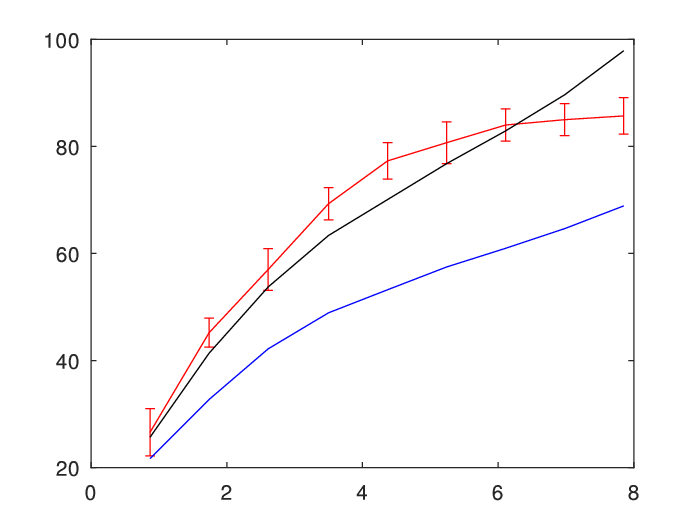}
         \caption{ UGC06667}
         \label{fig:UGC06667}
     \end{subfigure}
     \hfill
     \begin{subfigure}[b]{0.5\columnwidth}
         \centering
         \includegraphics[width=\columnwidth]{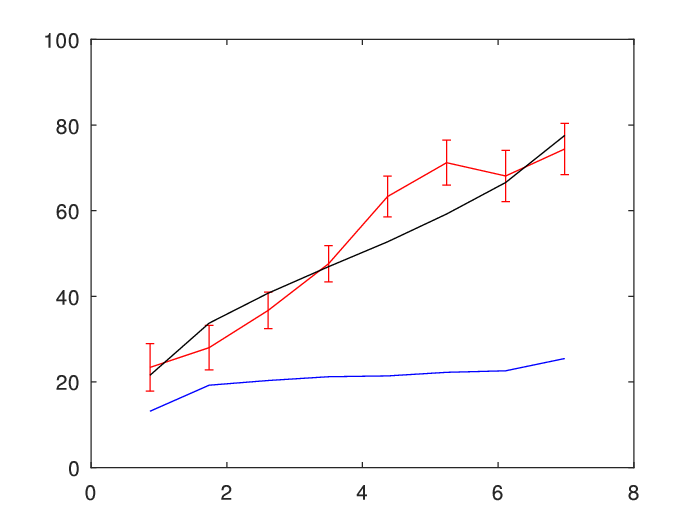}
         \caption{ UGC06818}
         \label{fig:UGC06818}
     \end{subfigure}
     \hfill
     \begin{subfigure}[b]{0.5\columnwidth}
         \centering
         \includegraphics[width=\columnwidth]{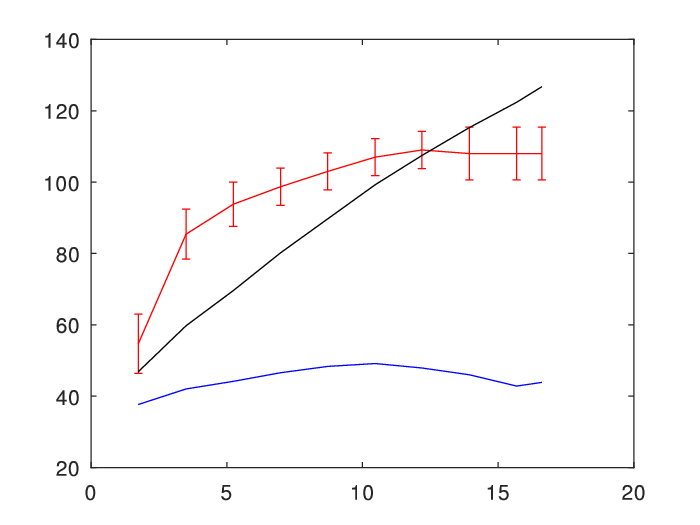}
         \caption{ UGC06930}
         \label{fig:UGC06930}
     \end{subfigure}
     \hfill
     \begin{subfigure}[b]{0.5\columnwidth}
         \centering
         \includegraphics[width=\columnwidth]{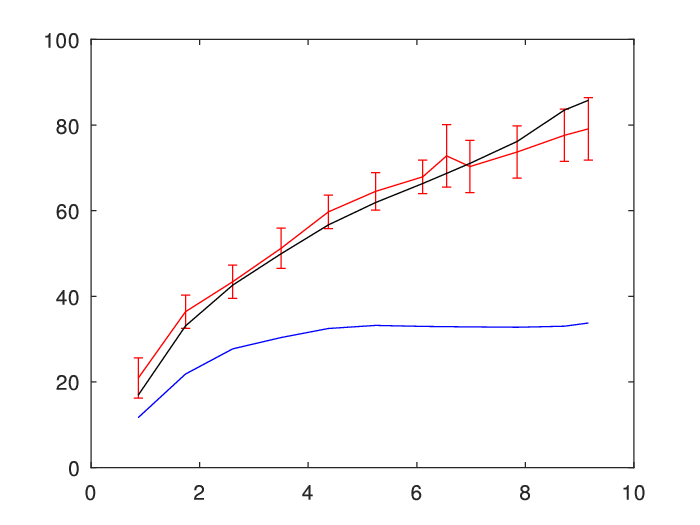}
         \caption{ UGC07089}
         \label{fig:UGC07089}
     \end{subfigure}
     \hfill
     \begin{subfigure}[b]{0.5\columnwidth}
         \centering
         \includegraphics[width=\columnwidth]{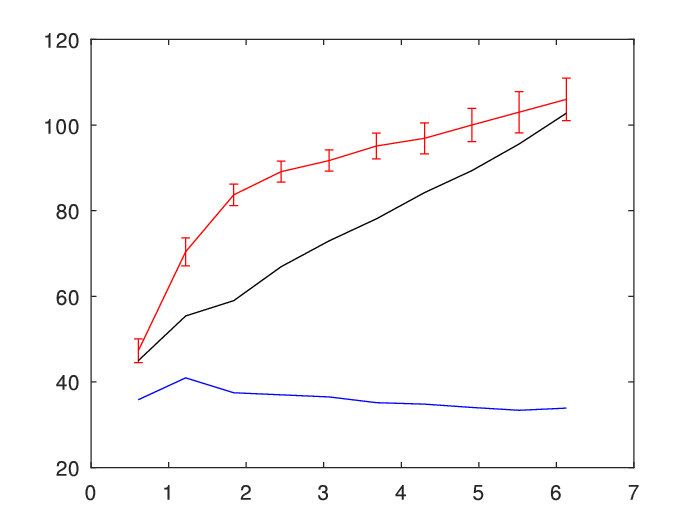}
         \caption{ UGC07399}
         \label{fig:UGC07399}
     \end{subfigure}
     \hfill
     \begin{subfigure}[b]{0.5\columnwidth}
         \centering
         \includegraphics[width=\columnwidth]{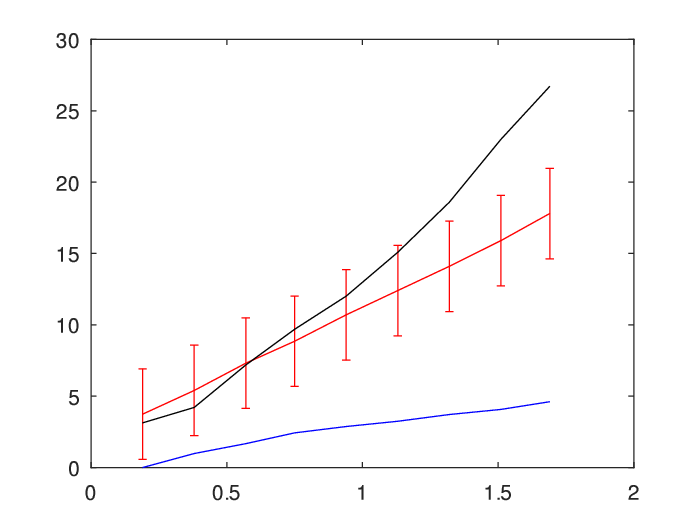}
         \caption{ UGC07577}
         \label{fig:UGC07577}
     \end{subfigure}
     \hfill
     \begin{subfigure}[b]{0.5\columnwidth}
         \centering
         \includegraphics[width=\columnwidth]{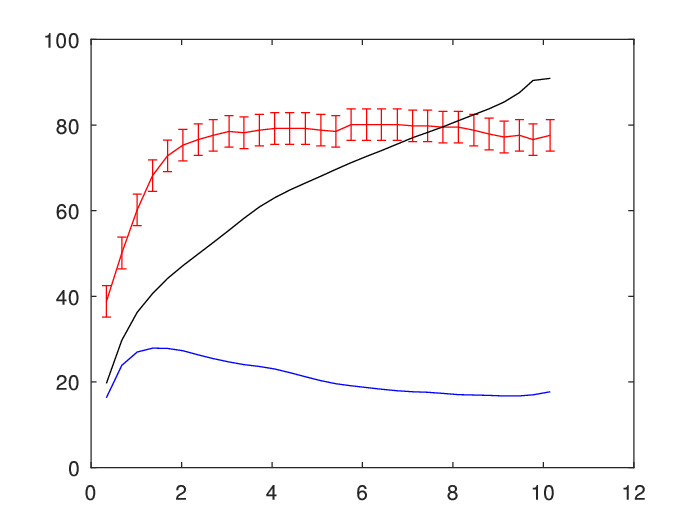}
         \caption{ UGC08490}
         \label{fig:UGC08490}
     \end{subfigure}
     \hfill
     \begin{subfigure}[b]{0.5\columnwidth}
         \centering
         \includegraphics[width=\columnwidth]{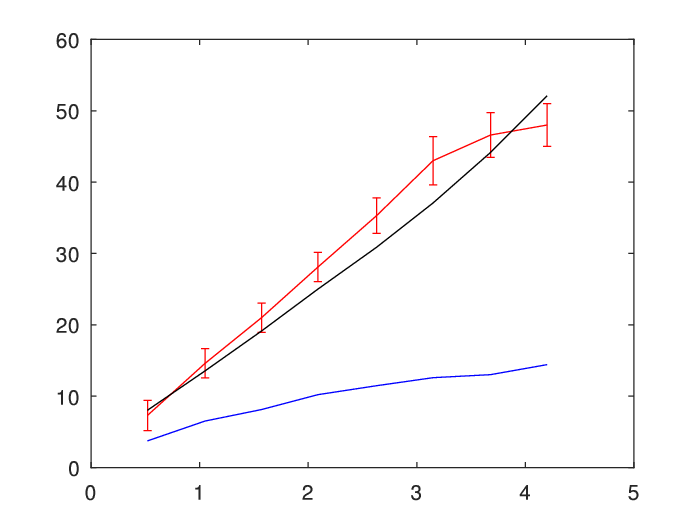}
         \caption{ UGC08837}
         \label{fig:UGC08837}
     \end{subfigure}
     \hfill
     \begin{subfigure}[b]{0.5\columnwidth}
         \centering
         \includegraphics[width=\columnwidth]{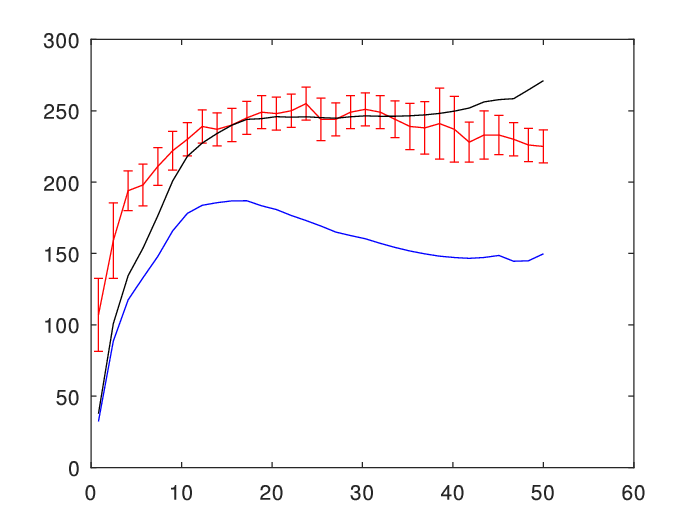}
         \caption{ UGC12506}
         \label{fig:UGC12506}
     \end{subfigure}
     \hfill
        \caption{Rotation Curves. The red curve is the experimentally observed curve, the black curve is calculated using MBG, and the blue curve is Newtonian.}
        \label{fig:sparc_rot_curves_c}
\end{figure*}

The details of the galaxies used for RAR simulation are provided in tables~\ref{tab:rar1} and~\ref{tab:rar2}. The corresponding figures are in Figs.~\ref{fig:sparc_rot_curves_a},~\ref{fig:sparc_rot_curves_b} and~\ref{fig:sparc_rot_curves_c}.
\begin{center}
\begin{table}
	\begin{tabular}{|c|c|c|c|}
\hline
 Galaxy & Galaxy mass & disk  & MBG r.m.s. err\\
 name  & x 1e9 $M_{\odot}$ & $\frac{M_{\odot}}{L_{\odot}}$ &(Newton r.m.s. err)\\
   & & & km/s \\
\hline
 CamB* & 0.21 & 1.2 & 1.60(22.46)\\
 F568-V1* & 6.40 & 2.2 & 26.45(57.67)\\
 F583-4 & 1.62 & 1.5 & 7.24(35.87)\\
 NGC0247 & 7.71 & 0.5 & 7.37(29.77)\\
 NGC2955 & 67.75 & 0.2 & 16.44(41.53)\\
 NGC3741* & 0.62 & 2.4 & 5.98(32.95)\\
 NGC3972 & 11.95 & 0.5 & 11.04(43.80)\\
 NGC4088 & 28.80 & 0.18 & 15.02(70.08)\\
 NGC5585* & 3.73 & 0.5 & 10.02(50.38)\\
 NGC6789 & 2.63 & 4.5 & 12.09(37.16)\\
 UGC00128* & 9.70 & 0.5 & 22.85(72.74)\\
 UGC03205 & 42.81 & 0.33 & 32.50(75.45)\\
 UGC04499* & 2.45 & 0.7 & 11.49(47.26)\\
 UGC05764* & 1.09 & 2.5 & 11.44(36.53)\\
 UGC06614 & 47.15 & 0.2 & 12.79(122.79)\\
 UGC06923 & 5.09 & 0.9 & 6.75(43.49)\\
 UGC07232 & 0.80 & 1.3 & 3.16(29.34)\\
 UGC07603 & 1.81 & 0.9 & 6.97(42.36)\\
 UGC08699 & 69.63 & 0.5 & 32.61(59.97)\\
 UGC11557 & 5.56 & 0.5 & 9.57(32.91)\\
 UGCA442* & 0.66 & 0.5 & 7.12(42.12)\\
 D631-7* & 0.60 & 0.7 & 3.86(38.25)\\
 DDO161* & 1.99 & 0.5 & 4.40(36.20)\\
 ESO079-G014 & 19.78 & 0.2 & 11.39(31.27)\\
 ESO563-G021 & 60.27 & 0.15 & 19.16(88.45)\\
 F574-1 & 7.81 & 1 & 13.62(37.93)\\
 F583-1* & 3.29 & 1.4 & 11.07(37.13)\\
 IC4202 & 33.77 & 0.1 & 27.96(59.15)\\
 NGC0055 & 6.50 & 0.2 & 11.52(30.43)\\
 NGC0289 & 49.90 & 0.3 & 24.68(70.28)\\
 NGC0891 & 61.80 & 0.25 & 15.68(93.37)\\
 NGC2683 & 27.93 & 0.15 & 31.67(81.51)\\
 NGC2915 & 2.83 & 0.7 & 9.79(65.86)\\
 NGC2998 & 43.83 & 0.3 & 21.93(71.48)\\
 NGC3521 & 83.26 & 0.5 & 21.34(62.44)\\
 NGC3769 & 14.29 & 0.23 & 22.21(61.43)\\
 NGC3917 & 12.43 & 0.4 & 8.89(38.42)\\
 NGC4013 & 23.52 & 0.08 & 24.05(105.39)\\
 NGC4085 & 18.01 & 0.3 & 10.82(56.63)\\
 NGC4214 & 2.85 & 0.5 & 11.84(60.71)\\
 NGC4559 & 12.98 & 0.35 & 11.03(55.87)\\
 NGC5055 & 44.52 & 0.2 & 23.63(71.97)\\
 NGC6015 & 19.66 & 0.5 & 16.06(67.16)\\
 NGC6674 & 64.25 & 0.3 & 19.13(98.49)\\
 NGC7331 & 76.62 & 0.16 & 15.64(102.55)\\
 UGC00891* & 1.05 & 0.6 & 5.60(43.94)\\
 UGC02487 & 100.15 & 0.5 & 10.32(126.23)\\
\hline
\end{tabular}
\caption{Galaxies from SPARC database used in RAR simulation. Galaxy names with a "*" have significant gas content.}
\label{tab:rar1}
\end{table}
\end{center}
\begin{center}
	\begin{table}
\begin{tabular}{|c|c|c|c|}
\hline
 Galaxy & Galaxy mass & disk  & MBG r.m.s. err\\
 name  & x 1e9 $M_{\odot}$ & $\frac{M_{\odot}}{L_{\odot}}$ &(Newton r.m.s. err)\\
   & & & km/s \\
\hline
 UGC02953 & 92.58 & 0.4 & 22.70(77.58)\\
 UGC03580 & 16.61 & 0.45 & 9.83(44.29)\\
 UGC04325 & 3.99 & 1.5 & 22.99(59.72)\\
 UGC05005* & 4.11 & 1 & 8.24(50.14)\\
 UGC05716* & 1.89 & 1.5 & 8.60(51.06)\\
 UGC05986 & 11.15 & 0.5 & 14.77(56.27)\\
 UGC06446* & 3.09 & 1.5 & 13.46(56.37)\\
 UGC06667 & 9.53 & 0.5 & 6.38(20.64)\\
 UGC06818 & 2.30 & 0.5 & 6.88(39.96)\\
 UGC06930 & 8.14 & 0.7 & 14.91(58.58)\\
 UGC07089 & 3.19 & 0.3 & 3.51(35.12)\\
 UGC07399 & 6.75 & 1.8 & 16.14(60.98)\\
 UGC07577 & 0.07 & 0.12 & 4.77(9.56)\\
 UGC08490* & 2.75 & 0.8 & 15.98(56.88)\\
 UGC08837 & 0.71 & 0.5 & 3.88(26.56)\\
 UGC12506 & 53.56 & 0.08 & 22.05(76.45)\\
\hline
\end{tabular}
\caption{More galaxies from SPARC database used in RAR simulation. Galaxy names with a "*" have significant gas content.}
\label{tab:rar2}
\end{table}
\end{center}
The r.m.s. errors provided in tables~\ref{tab:rar1} and~\ref{tab:rar2} are in the units of km/s, and estimates the r.m.s. error between the calculated rotation curve and the observed rotation curve.

\end{document}